\documentclass[12pt]{article}
\usepackage{caption}
\usepackage{amssymb}
\usepackage{amsmath}
\usepackage{hyperref}
\usepackage{float}
\usepackage{bm}
\usepackage{latexsym}
\usepackage{subfig}
\usepackage{mathtools}
\usepackage{booktabs}
\usepackage{tabularx} 
\usepackage{caption}    
\usepackage{subcaption} 
\usepackage{graphicx}
\usepackage{subcaption}
\usepackage{physics}
\usepackage{subfig}
\newcolumntype{C}{>{\centering\arraybackslash}X}
\title{\bf Thermodynamic and Topological Phase Transitions of AdS Black Holes with Nonminimal
$F^{\alpha\beta}F^{\gamma\lambda}R_{\alpha\gamma}R_{\beta\lambda}$ Coupling}
\author{Mehdi Sadeghi\thanks{Email: mehdi.sadeghi@abru.ac.ir}\,\,\, and \,\,Faramaz Rahmani\thanks{Corresponding author: faramarz.rahmani@abru.ac.ir}\hspace{2mm}\\
{\small {\em Department of Physics, Faculty of Basic Sciences,}}\\
{\small {\em Ayatollah Boroujerdi University, Boroujerd, Iran}}
}
\date{\today}
\begin{document}

\maketitle
\begin{abstract}
\noindent The conventional and topological phase transitions of a four-dimensional asymptotically AdS black hole with a non-minimal coupling term $F^{\alpha\beta}F^{\gamma\lambda}R_{\alpha\gamma}R_{\beta\lambda}$ are investigated. Using a perturbative approach to first order in the coupling $\epsilon$, the thermodynamic quantities are derived and the first law and Smarr relation are verified. Intriguingly, while the $\epsilon = 0$ limit yields the standard Reissner--Nordstr\"om--AdS black hole belonging to the $W^{1+}$ topological class ($W = +1$), switching on the non-minimal coupling fundamentally transforms the topology to the $W^{0-}$ class ($W = 0$). This transition occurs while the van der Waals-type first-order phase transition survives in the intermediate region, embedded within the overall Hawking--Page pattern, rendering the system a hybrid black hole thermodynamic system. The coupling $\epsilon$ thus acts as a topological deformation parameter that alters the universal classification of the system, despite the perturbative nature of the solution.
\end{abstract}


\section{Introduction} \label{intro}

General relativity, since its formulation by Einstein over a century ago, has remained the most successful geometric theory of gravity, passing a wide array of experimental tests from solar system dynamics to gravitational wave observations \cite{Wald1984, Carroll2004, Abbott2016}. Beyond its empirical success, General relativity fundamentally reshaped our understanding of spacetime, matter, and causality, establishing gravity as the curvature of spacetime and providing the backdrop for modern cosmology and black hole physics. 

A pivotal development in black hole physics was the discovery of their thermodynamic properties. The formulation of the four laws of black hole mechanics \cite{Bardeen:1973gs}, together with Hawking's seminal discovery of black hole radiation \cite{Hawking:1975vcx}, revealed that black holes are not merely geometric objects but also thermodynamic systems.
This analogy became particularly profound in asymptotically AdS spacetime, where the negative cosmological constant effectively acts as a natural confining box, allowing stable thermodynamic equilibrium configurations that are generally absent for isolated black holes in asymptotically flat spacetime. The work of Hawking and Page \cite{Hawking:1982dh} demonstrated a first-order phase transition between thermal AdS space and a large stable black hole. 
This Hawking-Page transition was later reinterpreted via the AdS/CFT correspondence \cite{ Witten:1998qj} as a confinement-deconfinement phase transition in the dual large-$N$ gauge theory living on the boundary of AdS.
 This holographic dictionary between gravitational thermodynamics in the bulk and quantum field theory thermodynamics on the boundary has become one of the most powerful tools for studying strongly coupled systems, including quark-gluon plasmas and unconventional superconductors \cite{Hartnoll:2009sz, Herzog:2009xv}.

The thermodynamic study of AdS black holes experienced a major development with the introduction of the extended phase-space framework  whereby  the cosmological constant $\Lambda$ is considered as a thermodynamic variable identified with the pressure, $P=-\frac{\Lambda}{8\pi G}$, whose conjugate quantity is the thermodynamic volume $V$. In this framework the black hole mass $M$ is no longer interpreted as internal  energy but rather as the enthalpy, satisfying the first law of thermodynamics~\cite{Kastor:2009wy}. This reinterpretation significantly enriched the phase structure of black holes, establishing a precise correspondence with classical fluid thermodynamics and enabling the identification of van der Waals-type phase transitions~\cite{Dolan2011,Kubiznak:2012wp}.

Earlier investigations had already shown that charged AdS black holes in the canonical ensemble undergo a first-order phase transition resembling the liquid-gas transition of a Van der Waals fluid \cite{Chamblin:1999tk,Chamblin:1999hg}. Within the extended phase-space picture, this analogy was placed on a firm thermodynamic basis through the formulation of an equation of state $P(V,T)$ and the identification of a critical point characterized by the mean-field critical exponents $(\alpha=0,\beta=1/2,\gamma=1,\delta=3)$ \cite{Dolan2011,Kubiznak:2012wp}. These developments established a deep correspondence between black-hole thermodynamics and classical fluid systems and stimulated extensive investigations of rich phenomena, including reentrant phase transitions \cite{Chemissany:2008fy,Kubiznak:2016qmn}, triple points and superfluid-like transitions \cite{Hennigar:2016xwd}. Since then, the extended phase-space formalism has been successfully generalized to rotating black holes \cite{Dolan:2013dga}, higher-curvature theories \cite{Cai:2013qga}, and black holes in arbitrary spacetime dimensions.

Given the richness of this thermodynamic landscape, the study of nonlinear gravitational theories in AdS spacetime has become an important direction in contemporary gravitational physics. Such non-linearities arise naturally from several well-motivated perspectives. First, higher-curvature corrections appear as the leading stringy corrections in low-energy effective action of string theory, with the Gauss-Bonnet term being a prominent example \cite{Zwiebach:1985uq}. Second, $f(R)$ gravity, where the Ricci scalar is replaced by a generic function $f(R)$, provides a simple and phenomenologically successful framework for modifying General Relativity at large scales \cite{Sotiriou:2008rp}. Third, scalar-tensor theories, including the archetypal Brans-Dicke theory \cite{BransDicke1961} and the more general Horndeski gravity which provides the most general framework for scalar-tensor modifications of gravity in four dimensions that remains free from Ostrogradsky instabilities, as it guarantees second-order equations of motion \cite{Horndeski:1974wa}. Fourth, non-minimal couplings between gauge fields and curvature, arise from one-loop quantum corrections in curved spacetime \cite{Drummond:1979pp} and from stringy scattering amplitudes \cite{Kats:2007mq}. Born–Infeld electrodynamics, which regularizes the divergent self-energy of point charges in Maxwell theory, represents another important non-linear modification of the gauge sector \cite{Balakin2010}. 
Understanding how these non-minimal and higher-curvature corrections shape black hole thermodynamics is important for several reasons. First, such terms may encode low-energy imprints of ultraviolet physics and provide insight into possible quantum-gravity effects. Second, within the AdS/CFT correspondence, higher-derivative interactions furnish valuable extensions of the holographic framework and influence the properties of the dual field theory. 

In recent years, a topological approach to black hole phase transitions has emerged, providing a classification scheme that is largely independent of the details of a particular equation of state. This framework is based on Duan's topological current $\phi$-mapping theory \cite{Duan1984,Duan1979}, originally developed in the study of topological defects in gauge field theory and condensed matter systems. 
In this approach, an off-shell free energy landscape is constructed for black holes. By defining a vector field $\phi$, whose first component is obtained from the derivative of the generalized free energy (which is proportional to  the Hawking--York Euclidean action~\cite{York:1986it}) with respect to the horizon radius, and whose second component is a function of an auxiliary angular parameter, each black hole solution is identified with a zero point of this vector field.
The topological charge, or winding number, $w$, associated with each zero point is determined by the degree of the mapping, which in two dimensions is expressed as a contour integral encircling the corresponding zero point. This construction provides a topological characterization of black hole equilibrium states and their phase transitions \cite{Wei:2021vdx,WeiLiu2026}. A notable feature of this framework is its topological robustness: the total topological number $W$, obtained by summing the winding numbers of all black-hole branches for a given set of thermodynamic parameters, is conserved under smooth deformations of the thermodynamic potentials.

Within this framework, each black-hole branch is represented by a topological defect, and the associated winding number $w_i=\pm1$ is generally correlated with its local thermodynamic stability: branches with positive heat capacity typically carry $w_i=+1$, whereas those with negative heat capacity are usually associated with $w_i=-1$. This perspective provides a topological characterization of the black-hole thermodynamic phase space and offers a defect-based interpretation of thermodynamic phase transitions.

The topological approach has since been extended to a wide variety of modified gravity theories, where numerous studies have shown that nonlinear coupling parameters can influence, and in some cases even alter, the topological class of the thermodynamic system \cite{Ali:2023jox,Gogoi:2025ied,Yerra:2022eov,EslamPanah:2024fls,Rahmani:2025iks,
Sadeghi:2026jic,Mehmood:2024nqd,Hazarika:2024xar}. These developments suggest that topological methods provide a unified framework for organizing and comparing black-hole thermodynamics across diverse gravitational theories. In particular, the total topological number \(W\) furnishes an additional characterization of the thermodynamic phase structure, playing a role analogous to that of topological invariants in condensed matter physics. Although the underlying physical systems are fundamentally different, both types of invariants encode global information that is insensitive to smooth deformations and thereby facilitate the classification of distinct phases.

The present work contributes to this growing body of research through a comprehensive topological investigation of an AdS black hole with the nonminimal interaction $
F^{\alpha\beta}F^{\gamma\lambda}R_{\alpha\gamma}R_{\beta\lambda}$. As shown in Sec.~\ref{sec2}, we derive a perturbative black-hole solution to first order in the coupling parameter \(\epsilon\), explicitly exhibiting the corrections induced by the nonminimal interaction to the metric function, gauge field, and thermodynamic quantities. In Sec.~\ref{sec3}, we investigate the thermodynamic properties of the system in the canonical ensemble by analyzing the temperature, entropy, Helmholtz free energy, and heat capacity, thereby elucidating how the coupling parameter \(\epsilon\) modifies the critical behavior of the system. In Sec.~\ref{sec4}, we introduce the topological framework based on the off-shell free energy and the associated vector field, and define the topological charge within Duan's topological current theory. Finally, in Sec.~\ref{sec5}, we analyze the topological phase structure of the nonminimally coupled black hole and show that the coupling parameter \(\epsilon\) can act as a topological deformation parameter, inducing transitions between distinct winding-number classes.

\section{Nonminimally Coupled AdS Black Hole and Perturbative Solutions}\label{sec2}

The bulk action that defines our model is
\begin{equation}\label{action}
	S = \int d^{4} x \sqrt{-g} \bigg[ \frac{1}{\kappa}(R - 2\Lambda) - \tfrac{1}{4} \alpha  F_{\mu \nu }F^{ \mu \nu  } + \epsilon F^{\alpha \beta } F^{\gamma \lambda } R_{\alpha \gamma } R_{\beta \lambda }  \bigg],
\end{equation}
where \(R\) is the Ricci scalar, \(\Lambda = -3/l^{2}\) is the cosmological constant (with \(l\) the AdS radius), and \(\kappa = 16\pi G\) in standard units. The Maxwell strength tensor field is  defined as
\begin{align} \label{YM}
	F_{\mu \nu } =\partial _{\mu } A_{\nu } -\partial _{\nu } A_{\mu } ,
\end{align}
and \(A_{\mu}\) the gauge potential. The Ricci tensor is denoted by \(R_{\mu\nu}\).

The parameter $\epsilon$ controls the strength of the nonminimal interaction between the electromagnetic field and spacetime curvature and has dimensions of $L^4$. Such higher-dimensional interactions are naturally motivated within the framework of effective field theory, where gauge fields and spacetime curvature are expected to couple through higher-order operators generated by quantum corrections or obtained after integrating out heavy degrees of freedom. Similar nonminimal gauge-curvature couplings also arise in low-energy effective actions inspired by string theory and other ultraviolet completions of gravity. The interaction considered here belongs to this class of higher-dimensional operators and provides a simple phenomenological framework for investigating how nonlinear matter-geometry couplings influence black-hole solutions and their thermodynamic properties. From a phenomenological perspective, such interactions offer a controlled setting for exploring deviations from Einstein-Maxwell theory and for examining the effects of higher-curvature corrections on black-hole geometry, phase structure, and  critical behavior\cite{Balakin:2005fu,Balakin2010}.

To construct static, spherically symmetric black-hole solutions, we consider the metric ansatz
\begin{equation}\label{metric}
ds^{2}=-f(r)e^{-2H(r)}dt^{2}
+\frac{dr^{2}}{f(r)}
+r^{2}\left(d\theta^{2}+\sin^{2}\theta\,d\phi^{2}\right),
\end{equation}
where the redshift function \(H(r)\) encodes the effects of the nonminimal matter--curvature interaction. For a purely electric configuration, the gauge potential is taken to be
\begin{equation}\label{background}
A_{\mu}dx^{\mu}=h(r)\,dt,
\end{equation}
which gives rise to the field-strength tensor
\begin{equation}
\label{YM1}
F_{\mu\nu}
=
\left(
\begin{array}{cccc}
0 & -h'(r) & 0 & 0\\
h'(r) & 0 & 0 & 0\\
0 & 0 & 0 & 0\\
0 & 0 & 0 & 0
\end{array}
\right).
\end{equation}
Consequently, the Maxwell invariant becomes
\begin{equation}
\mathcal{F}
=
F_{\mu\nu}F^{\mu\nu}
=
-2e^{2H(r)}h'(r)^2.
\end{equation}

Variation of the action (\ref{action}) with respect to the metric yields the gravitational field equations
\begin{equation}\label{EOM1}
R_{\mu\nu}
-\frac12g_{\mu\nu}R
+\Lambda g_{\mu\nu}
=
\kappa T_{\mu\nu}^{(\mathrm{eff})},
\end{equation}
where
\begin{equation}
T_{\mu\nu}^{(\mathrm{eff})}
=
\alpha T_{\mu\nu}^{(\mathrm{M})}
+
\epsilon T_{\mu\nu}^{(\mathrm{I})}.
\end{equation}
Here,
\begin{equation}
T_{\mu\nu}^{(\mathrm{M})}
=
\frac12F_{\mu}{}^{\alpha}F_{\nu\alpha}
-
\frac18g_{\mu\nu}
F_{\alpha\beta}F^{\alpha\beta}
\end{equation}
is the standard Maxwell energy-momentum tensor, whereas \(T_{\mu\nu}^{(\mathrm{I})}\) denotes the contribution arising from the nonminimal interaction. Owing to its length, the explicit form of \(T_{\mu\nu}^{(\mathrm{I})}\) is presented in Appendix~A.

The Maxwell equations obtained from the action are
\begin{equation}\label{EOM-YM}
\nabla_{\nu}\left(
-\frac{\alpha}{2}F^{\mu\nu}
+
2\epsilon
F^{\alpha\beta}
R_{\alpha}{}^{\mu}
R_{\beta}{}^{\nu}
\right)=0.
\end{equation}

Because of the nonlinear nature of the nonminimal interaction, obtaining an exact analytical solution appears to be intractable. We therefore employ a perturbative expansion in the small coupling parameter \(\epsilon\). Accordingly, the metric functions and gauge potential are expanded as
\begin{align}
f(r)&=f_0(r)+\epsilon f_1(r)+\mathcal{O}(\epsilon^2),\\
H(r)&=H_0(r)+\epsilon H_1(r)+\mathcal{O}(\epsilon^2),\\
h(r)&=h_0(r)+\epsilon h_1(r)+\mathcal{O}(\epsilon^2).
\end{align}
At zeroth order, corresponding to \(\epsilon=0\), the theory reduces to Einstein--Maxwell gravity. The \(tt\) and \(rr\) components of the field equations (Apeendix B) at this order become
\begin{align}
4rf_0'+4f_0+\kappa\alpha r^2e^{2H_0}h_0'^2
+4\Lambda r^2&=0,\\
4rf_0'+4f_0-8rf_0H_0'
+\kappa\alpha r^2e^{2H_0}h_0'^2
+4\Lambda r^2&=0.
\end{align}
Subtracting these equations yields \(H_0'(r)=0\). The corresponding integration constant can be absorbed by a rescaling of the time coordinate, allowing one to set
\begin{equation}
H_0(r)=0.
\end{equation}
The remaining equations then give the familiar Reissner--Nordstr\"om--AdS solution,
\begin{equation}\label{f0}
f_0(r)=1-\frac{2m_0}{r}
-\frac{\Lambda r^2}{3}+\frac{\alpha\kappa Q^2}{4r^2},
\qquad
m_0=-\frac{\Lambda r_h^3}{6}+\frac{\kappa\alpha Q^2}{8r_h}
+\frac{r_h}{2},
\end{equation}
together with
\begin{equation}\label{h0}
h_0(r)=Q\left(\frac{1}{r}-\frac{1}{r_h}\right),
\end{equation}
where \(r_h\) denotes the radius of the outer event horizon.

Proceeding to first order in the coupling parameter \(\epsilon\), we substitute the perturbative expansions into the complete field equations. Subtracting the \(tt\) and \(rr\) components of the Einstein field equations at \(\mathcal{O}(\epsilon)\) yields
\begin{equation}
64\kappa r f_0 h_0' h_0'' + 32\kappa r^2 f_0 h_0^{(3)} h_0' + 32\kappa r^2 f_0 (h_0'')^2 - 8r f_0 H_1' = 0.
\end{equation}
Integrating this equation gives
\begin{equation}\label{H1}
H_1(r)=C_1-\frac{3}{2}\kappa f_0''h_0'^2
-2\kappa f_0'h_0'h_0''
-\kappa r f_0''h_0'h_0''
-\frac12\kappa r f_0^{(3)}h_0'^2.
\end{equation}
Substituting the zeroth-order solutions (\ref{f0}) and (\ref{h0}) into Eq.~(\ref{H1}), one finds
\begin{equation}\label{HH}
H_1(r)
=
C_1
-\frac{3\Lambda\kappa Q^2}{r^4}
+\frac{7\kappa^2\alpha Q^4}{4r^8}.
\end{equation}
The integration constant \(C_1\) is fixed by requiring the boundary speed of light to be unity, which implies \(C_1=0\) \cite{Balakin2010}.

At first order in \(\epsilon\), the Maxwell equation yields
\begin{equation}\label{h1}
h_1(r)
=
-\frac{\kappa^2\alpha Q^5}{6r^9}
+\frac{\kappa Q^3\Lambda}{5r^5}
+\frac{4\Lambda^2Q}{\alpha r}
+C_2.
\end{equation}

The \(rr\) component of the Einstein field equations at \(\mathcal{O}(\epsilon)\) determines the metric correction,
\begin{align}\label{f1}
f_1(r)
&=
\frac{109\alpha^2\kappa^3Q^6}{144r^{10}}
-\frac{13m_0\alpha\kappa^2Q^4}{2r^9}
+\frac{7\alpha\kappa^2Q^4}{2r^8}
\nonumber\\
&
\quad
-\frac{73\alpha\kappa^2Q^4\Lambda}{30r^6}
+\frac{10m_0\kappa Q^2\Lambda}{r^5}
-\frac{6\kappa Q^2\Lambda}{r^4}
+\frac{11\kappa Q^2\Lambda^2}{3r^2}
-\frac{C_3}{r}.
\end{align}

To determine the remaining integration constants and maintain consistency of the perturbative expansion, we introduce a perturbed horizon radius,
\begin{equation}
r_h'=r_h+\epsilon r_h^{(1)},
\end{equation}
where \(r_h\) denotes the unperturbed horizon radius and \(\epsilon r_h^{(1)}\) its first-order correction.

Expanding the condition \(h(r_h')=0\) gives
\begin{align}
h(r_h+\epsilon r_h^{(1)})
=
h_0(r_h)
+
\epsilon
\left[
h_0'(r_h)r_h^{(1)}
+h_1(r_h)
\right]
+
\mathcal O(\epsilon^2)
=
0,
\end{align}
which leads to
\begin{equation}
r_h^{(1)}
=
-\frac{h_1(r_h)}{h_0'(r_h)}.
\end{equation}
Hence, the perturbed horizon radius becomes
\begin{equation}\label{ph}
r_h'
=
r_h
-
\left(
\frac{h_1(r_h)}{h_0'(r_h)}
\right)\epsilon.
\end{equation}

Imposing the horizon condition \(h(r_h')=0\) at first order fixes
\begin{equation}
C_2
=
-\frac{4\Lambda^2Q}{\alpha r_h}
-\frac{\kappa Q^3\Lambda}{5r_h^5}
+
\frac{\kappa^2\alpha Q^5}{6r_h^9}.
\end{equation}

For consistency of the perturbative description, the horizon displacement inferred from the gauge potential should coincide with that obtained from the metric function, leading to the matching condition
\begin{equation}\label{con}
\frac{h_1(r_h)}{h_0'(r_h)}
=
\frac{f_1(r_h)}{f_0'(r_h)}.
\end{equation}

Since, \(h_1(r_h)=0\), Eq.~(\ref{con}) further requires \(f_1(r_h)=0\), thereby fixing the integration constant \(C_3\) as
\begin{equation}
C_3
=
-\frac{2\kappa Q^2\Lambda^2}{r_h}
+
\frac{\kappa^2Q^4\Lambda\alpha}{10r_h^5}
+
\frac{\kappa^3Q^6\alpha^2}{18r_h^9}
+
\frac{\kappa Q^2\Lambda}{r_h^3}
-
\frac{\kappa^2Q^4\alpha}{4r_h^7}.
\end{equation}

Collecting the above results, the gauge potential and metric function up to first order in $\epsilon$ are obtained as
\begin{equation}\label{hfinal}
\begin{aligned}
h(r) = \frac{Q}{r} - \frac{Q}{r_h} 
+ \epsilon \Bigg[
&-\frac{\kappa^2 \alpha Q^5}{6 r^9}
+ \frac{\kappa Q^3 \Lambda}{5 r^5}
+ \frac{4 \Lambda^2 Q}{\alpha r} \\
&- \frac{4 \Lambda^2 Q}{\alpha r_h}
- \frac{\kappa Q^3 \Lambda}{5 r_h^5}
+ \frac{\kappa^2 \alpha Q^5}{6 r_h^9}
\Bigg],
\end{aligned}
\end{equation}
and
\begin{equation}\label{f_metric}
\begin{aligned}
f(r) = &\; 1 - \frac{\Lambda r^2}{3} - \frac{2M}{r} 
+ \frac{\kappa Q^2}{r^2}\left(\frac{\alpha}{4} + \frac{11}{3}\epsilon \Lambda^2\right)
- \frac{6\epsilon \kappa Q^2 \Lambda}{r^4} \\
&+ \frac{5\epsilon \kappa Q^2 \Lambda}{r^5}\left(r_h - \frac{\Lambda r_h^3}{3} + \frac{\kappa \alpha Q^2}{4 r_h}\right)
- \frac{73\epsilon \alpha \kappa^2 Q^4 \Lambda}{30 r^6} \\
&+ \frac{7\epsilon \alpha \kappa^2 Q^4}{2 r^8}
+ \frac{13\epsilon \kappa^2 Q^4 \alpha}{r^9}\left(\frac{\Lambda r_h^3}{12} - \frac{r_h}{4} - \frac{\kappa \alpha Q^2}{16 r_h}\right)
+ \frac{109 \epsilon \alpha^2 \kappa^3 Q^6}{144 r^{10}}.
\end{aligned}
\end{equation}

The corresponding mass parameter is
\begin{equation}\label{ADM}
\begin{aligned}
M
&=
\underbrace{
\left(
-\frac{\Lambda r_h^3}{6}
+\frac{r_h}{2}
+\frac{\kappa\alpha Q^2}{8r_h}
\right)
}_{m_0}
\\
&\quad
+
\epsilon
\underbrace{
\left(
\frac{\kappa Q^2\Lambda^2}{r_h}
-
\frac{\kappa Q^2\Lambda}{2r_h^3}
-
\frac{\kappa^2Q^4\alpha\Lambda}{20r_h^5}
+
\frac{\kappa^2Q^4\alpha}{8r_h^7}
-
\frac{\kappa^3Q^6\alpha^2}{36r_h^9}
\right)
}_{m_1}.
\end{aligned}
\end{equation}

To establish the physical interpretation of \(M\), we examine the asymptotic behavior of
\(g_{tt}=-f(r)e^{-2H(r)}\).
Using the perturbative expansions \(f=f_0+\epsilon f_1\) and \(H=\epsilon H_1\), one finds
\[
g_{tt}
=
-\left(
1-\frac{2m_0}{r}
-\frac{\Lambda r^2}{3}
+\frac{\kappa\alpha Q^2}{4r^2}
\right)
-\epsilon
\bigl(
f_1(r)-2f_0(r)H_1(r)
\bigr)
+\mathcal O(\epsilon^2).
\]
Substituting Eqs.~(\ref{f1}) and (\ref{HH}), one obtains
\[
f_1(r)-2f_0(r)H_1(r)
=
-\frac{2m_1}{r}
+
\mathcal O\!\left(\frac1{r^4}\right),
\]
Consequently,
\[
g_{tt}
=
-
\left(
1
-
\frac{\Lambda r^2}{3}
-
\frac{2(m_0+\epsilon m_1)}{r}
\right)
+
\mathcal O\!\left(\frac1{r^4}\right).
\]

Therefore, the conserved mass parameter of the black hole to first order in \(\epsilon\) is identified as
\[
M=m_0+\epsilon m_1,
\]
in agreement with the first law of black-hole thermodynamics, as will be demonstrated in Sec.~\ref{sec3}.


\section{Thermodynamic Phase Structure in the Canonical Ensemble}\label{sec3}

In this section, we study the thermodynamic behavior of the system in the canonical ensemble. Although the criticality and thermodynamic properties of the system can be studied comprehensively and in great detail, we restrict our analysis to those aspects that are essential for the interpretation and implementation of the final part of our work, namely the topological investigation.

\subsection{Essential Thermodynamic Relations and the First Law of Thermodynamics}
To study the thermodynamics of the system in the extended phase space, we adopt the standard relation $\Lambda = -8\pi G P$ and set $G=1$ for simplicity, which allows us to interpret the ADM mass as the enthalpy. The Hawking temperature is obtained separately from the surface gravity formula. From this point onward, we set $\kappa = 1$ and $\alpha = 1$ in all calculations. To avoid confusion with the redshift function $H(r)$, we denote the enthalpy by $\mathcal{H}$, which reads
\begin{equation}\label{enthalpy}
\mathcal{H} = \left( \frac{r_h}{2} + \frac{Q^2}{8 r_h} + \frac{4 \pi P r_h^3}{3} \right) + \left( -\frac{Q^6}{36 r_h^9} + \frac{Q^4}{8 r_h^7} + \frac{2 \pi P Q^4}{5 r_h^5} + \frac{4 \pi P Q^2}{r_h^3} + \frac{64 \pi^2 P^2 Q^2}{r_h} \right) \epsilon.
\end{equation}

The Hawking temperature is computed via the surface gravity formula,
\begin{equation}\label{Temp}
\begin{split}
T &= \frac{1}{2\pi} \left[ \frac{1}{\sqrt{g_{rr}}} \frac{d}{dr} \sqrt{-g_{tt}} \right] \Bigg|_{r = r_h} 
   = \frac{e^{-H(r_h)} f'(r_h)}{4 \pi} \\
  &=  2Pr_h + \frac{1}{4\pi r_h} - \frac{Q^2}{16\pi r_h^3} + \left( -\frac{2PQ^2}{r_h^5} + \frac{PQ^4}{r_h^7} - \frac{Q^4}{16\pi r_h^9} + \frac{Q^6}{32\pi r_h^{11}} \right) \epsilon.
\end{split}
\end{equation}

The black hole entropy follows from the thermodynamic identity
\begin{equation}
S = \int \frac{1}{T} \frac{\partial \mathcal{H}}{\partial r_h} \, dr_h 
    = \pi r_h^2 + \left( \frac{16\pi^2 P Q^2}{r_h^2} + \frac{\pi Q^4}{2 r_h^6} \right) \epsilon.
\end{equation}

Using Eq.~\eqref{Temp}, the pressure can be expressed in terms of the temperature and horizon radius as
\begin{equation}\label{pre}
P = \frac{32\pi T r_h^{11} + 2Q^2 r_h^8 - 8 r_h^{10} - Q^6 \epsilon + 2Q^4 \epsilon r_h^2}{32\pi r_h^4 \left( 2 r_h^8 + Q^4 \epsilon - 2 Q^2 \epsilon r_h^2 \right)}.
\end{equation}

The denominator in these expressions may vanish for certain parameter values, potentially indicating singularities. Importantly, the critical points (Table \ref{tab:critical_points_selected}), where phase transitions occur, and the physically admissible parameter regime lie well away from these singularities, ensuring that all thermodynamic quantities remain well-behaved in the region of interest.

Expanding the pressure as a series in $\epsilon$ gives
\begin{equation}\label{pre2}
P=\frac{T}{2r_h} - \frac{1}{8\pi r_h^2} + \frac{Q^2}{32\pi r_h^4} + \left( \frac{Q^2 T}{2 r_h^7} - \frac{Q^4 T}{4 r_h^9} - \frac{Q^2}{8\pi r_h^8} + \frac{Q^4}{8\pi r_h^{10}} - \frac{Q^6}{32\pi r_h^{12}} \right) \epsilon.
\end{equation}

To have a consistent Smarr relation, the first law of thermodynamics in the extended phase space takes the form
\begin{equation}\label{first}
d\mathcal{H} = T\,dS + V\,dP + \psi\,dQ + \Xi\,d\epsilon,
\end{equation}
where $T$, $V$, $\psi$, and $\Xi$ are the thermodynamic conjugates to the entropy $S$, pressure $P$, charge $Q$, and the nonminimal coupling parameter $\epsilon$, respectively.

The temperature is obtained from the surface gravity formula and, equivalently, from the thermodynamic identity $T = (d\mathcal{H}/dr_h)/(dS/dr_h)$, yielding
\begin{equation}\label{temp2}
\begin{aligned}
T 
&= 2Pr_h + \frac{1}{4\pi r_h} - \frac{Q^2}{16\pi r_h^3} 
 + \left( -\frac{2PQ^2}{r_h^5} + \frac{PQ^4}{r_h^7} - \frac{Q^4}{16\pi r_h^9} + \frac{Q^6}{32\pi r_h^{11}} \right) \epsilon.
\end{aligned}
\end{equation}

The thermodynamic volume, defined as the Legendre-transform quantity conjugate to the pressure, is
\begin{equation}\label{volume}
\begin{aligned}
V 
&= \left( \frac{\partial \mathcal{H}}{\partial P} \right)_{Q, S} \\
&= \left( \frac{\partial \mathcal{H}}{\partial P} \right)_{Q, r_h} 
   - \frac{
       \left( \frac{\partial \mathcal{H}}{\partial r_h} \right)_{P,Q}
       \left( \frac{\partial S}{\partial P} \right)_{Q, r_h}
     }{
       \left( \frac{\partial S}{\partial r_h} \right)_{P,Q}
     } 
= \frac{4\pi r_h^3}{3} 
   + \epsilon \left( \frac{96\pi^2 P Q^2}{r_h} + \frac{7\pi Q^4}{5 r_h^5} \right),
\end{aligned}
\end{equation}
while the electric potential, conjugate to the charge, reads
\begin{equation}\label{potential}
\begin{aligned}
\psi 
&= \left( \frac{\partial \mathcal{H}}{\partial Q} \right)_{P, S} \\
&= \left( \frac{\partial \mathcal{H}}{\partial Q} \right)_{P, r_h} 
   - \frac{
       \left( \frac{\partial \mathcal{H}}{\partial r_h} \right)_{P,Q}
       \left( \frac{\partial S}{\partial Q} \right)_{P, r_h}
     }{
       \left( \frac{\partial S}{\partial r_h} \right)_{P,Q}
     } 
= \frac{Q}{4r_h} 
   + \epsilon \left( \frac{64\pi^2 P^2 Q}{r_h} - \frac{2\pi P Q^3}{5 r_h^5} - \frac{Q^5}{24 r_h^9} \right).
\end{aligned}
\end{equation}

The conjugate potential $\Xi$ to the perturbation coefficient $\epsilon$ is
\begin{equation}\label{Xi}
\begin{aligned}
\Xi 
&= \left( \frac{\partial \mathcal{H}}{\partial \epsilon} \right)_{S,P,Q} 
= \left(\pdv{\mathcal{H}}{\epsilon}\right)_{P,Q,r_h} 
   - \frac{
      \left( \pdv{\mathcal{H}}{r_h}\right)_{P,Q,\epsilon}
        \left(\pdv{S}{\epsilon}\right)_{P,Q,r_h}
     }{
       \left(\pdv{S}{r_h}\right)_{P,Q,\epsilon}
     } \\ \\
&= \frac{32\pi^2P^2Q^2}{r_h}
   + \frac{2\pi P Q^4}{5r_h^5}
   + \frac{Q^6}{288r_h^9} \\
&\quad + \epsilon
   \left(
     \frac{256\pi^3P^2Q^4}{r_h^5}
     + \frac{64\pi^2P Q^6}{5r_h^9}
     + \frac{11\pi Q^8}{72r_h^{13}}
   \right) \\
&= \Xi_0 + \epsilon \Xi_1.
\end{aligned}
\end{equation}

The expressions above have been computed perturbatively to first order in the non-minimal coupling parameter $\epsilon$, with all results truncated at $\mathcal{O}(\epsilon)$.

With these relations, the Smarr relation up to first order in $\epsilon$ takes the form
\begin{equation}\label{Smarr}
\mathcal{H} = 2TS - 2PV + \psi Q + 4\Xi_0 \epsilon.
\end{equation}
This relation confirms the correctness of the modified mass parameter $M$ in the nonminimal model.

\subsection{Critical Behavior and Thermodynamic Phase Structure}
It is now appropriate to verify whether the temperature and pressure satisfy the criticality conditions

\begin{equation}\label{crit_p}
\left(\frac{\partial P}{\partial r_h}\right)_T = \left(\frac{\partial^2 P}{\partial r_h^2}\right)_T = 0,
\end{equation}

or, equivalently,

\begin{equation}\label{crit_t}
\left(\frac{\partial T}{\partial r_h}\right)_P = \left(\frac{\partial^2 T}{\partial r_h^2}\right)_P = 0.
\end{equation}

To this end, we investigate this behavior for a fixed value of the non-minimal coupling parameter, \(\epsilon = 0.001\).
 As demonstrated in Appendix~\ref{app:epsilon}, this choice lies well within the regime where the perturbative expansion is reliable; the appendix also provides a detailed discussion of the validity range of \(\epsilon\). To plot and interpret the results, we employ the critical data summarized in Table~\ref{tab:critical_points_selected}.

\begin{table}[H]
\centering
\caption{Critical points for selected values of the electric charge $Q$ with fixed $\epsilon = 0.001$, $\kappa = 1$, and $\alpha = 1$.}
\label{tab:critical_points_selected}
\begin{tabularx}{\textwidth}{C C C C}
\toprule
$Q$ & $r_c$ & $T_c$ & $P_c$ \\
\midrule
0.00 & \multicolumn{3}{c}{No critical point} \\
\hline
0.01 & 0.02754066 & 2.22411651 & 1178.08599000 \\
0.02 & 0.05533406 & 1.11567180 & 18.61392410 \\
0.05 & 0.15855795 & 0.69089202 & 0.80188168 \\
0.08 & 0.18968511 & 0.60940499 & 0.62465653 \\
0.10 & 0.21212796 & 0.54922338 & 0.50906985 \\
0.20 & 0.30514803 & 0.37133448 & 0.23765128 \\
0.30 & 0.39977038 & 0.27364375 & 0.13092489 \\
0.40 & 0.50659385 & 0.21216848 & 0.07920185 \\
0.50 & 0.62154053 & 0.17171205 & 0.05200220 \\
0.60 & 0.74029964 & 0.14374600 & 0.03647858 \\
0.70 & 0.86079708 & 0.12346049 & 0.02692122 \\
0.80 & 0.98213828 & 0.10813593 & 0.02065734 \\
0.90 & 1.10392080 & 0.09617240 & 0.01634131 \\
1.00 & 1.22595024 & 0.08658183 & 0.01324552 \\
\bottomrule
\end{tabularx}
\end{table}

\begin{figure}[H]
    \centering
    \includegraphics[width=0.38\textwidth]{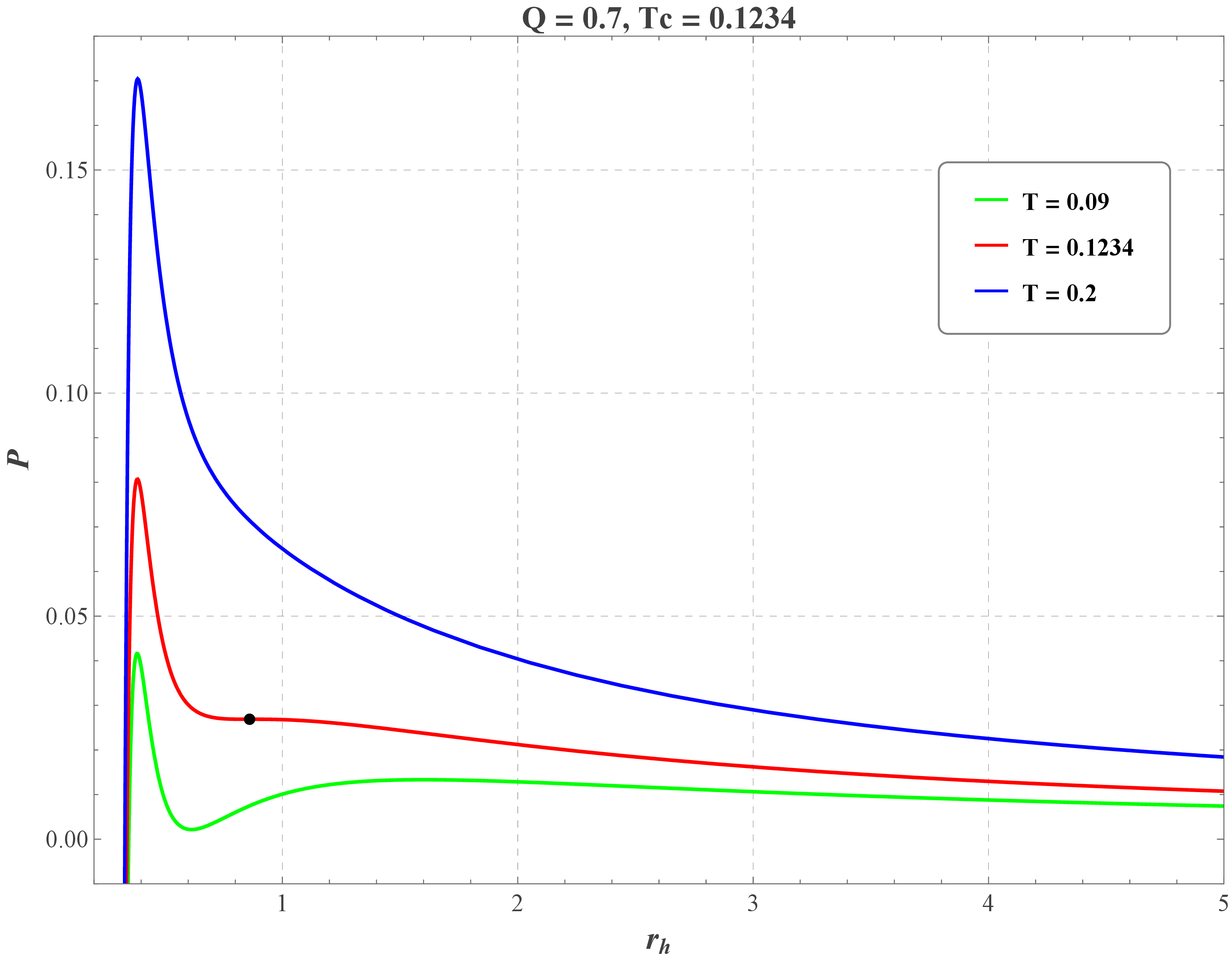}
    \quad
    \includegraphics[width=0.38\textwidth]{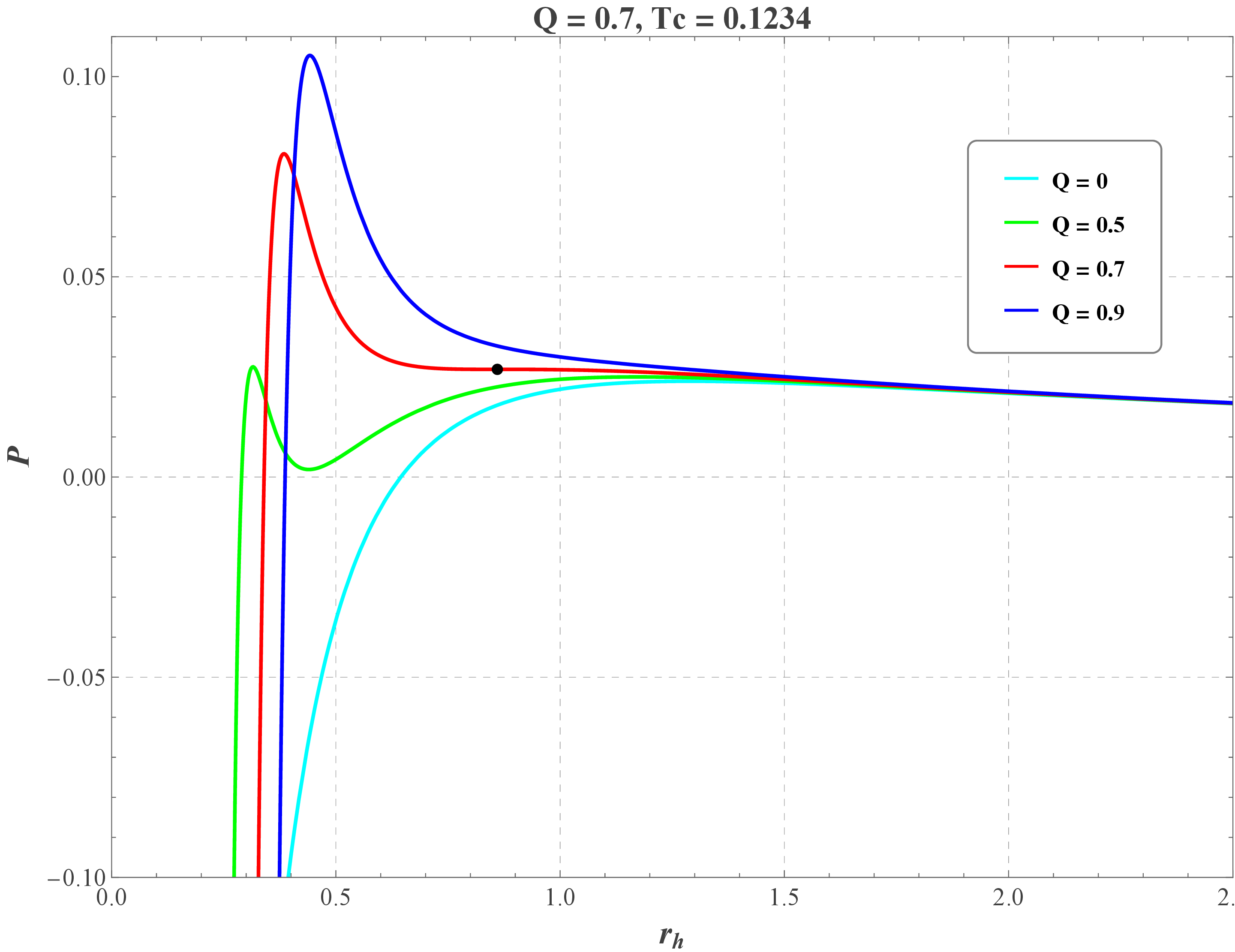}
    \caption{(a) Isothermal $P$--$r_h$ diagrams for fixed $Q = 0.7$ at temperatures above, at, and below the critical temperature $T_c = 0.1234$. At small horizon radii, the behavior resembles that of a Hawking--Page system, while below $T_c$, a characteristic van der Waals oscillation appears in the intermediate region. (b) Pressure as a function of horizon radius for different values of the electric charge $Q$ at fixed temperature. For $Q = 0$, the system reduces to the pure Hawking--Page case, whereas nonzero $Q$ introduces deviations that become more pronounced as $Q$ increases.}
    \label{fig:pressure}
\end{figure}

The pressure curves displayed in Fig.~\ref{fig:pressure} clearly illustrate the critical behavior. In the limit $r_h \to 0$, the pressure does not diverge to positive infinity as in a pure van der Waals system; instead, it exhibits a Hawking--Page-like behavior, suggesting a more intricate underlying structure. 

The left panel shows the isotherm diagram for $Q = 0.7$ at the critical temperature $T_c = 0.1234$. Above $T_c$, the isotherms are smooth, while below $T_c$, a characteristic van der Waals oscillation emerges. The right panel illustrates the pressure behavior as $Q$ varies. For $Q = 0$, the system reduces to the pure Hawking--Page case. For $P < P_c$ and sufficiently large $Q$, the isotherm develops a small-loop oscillation (see Fig.~\ref{fig:pressure}), signaling a VdW-type first-order phase transition between small and intermediate black holes. 

\begin{figure}[H]
    \centering
    \includegraphics[width=0.38\textwidth]{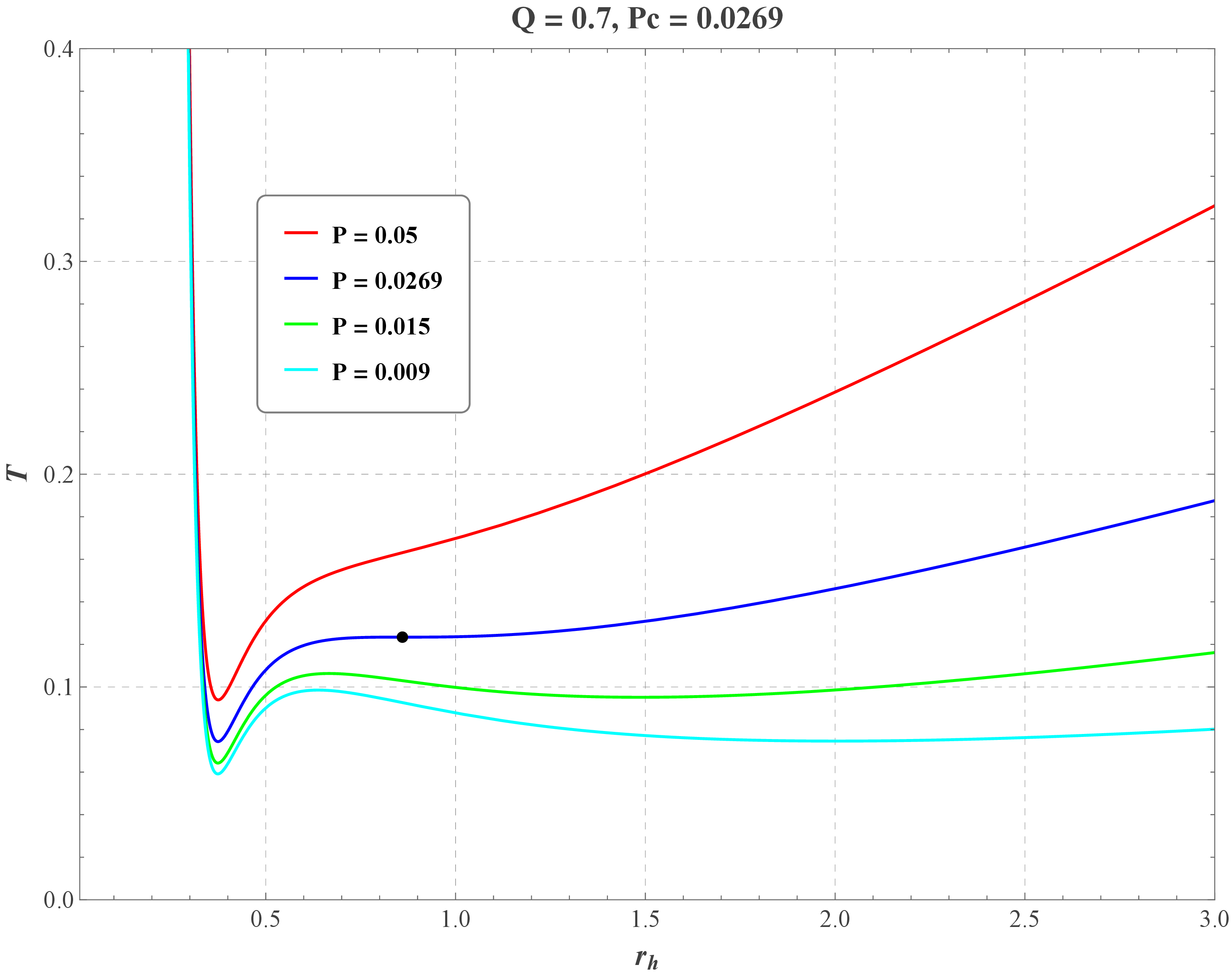}
    \quad
    \includegraphics[width=0.38\textwidth]{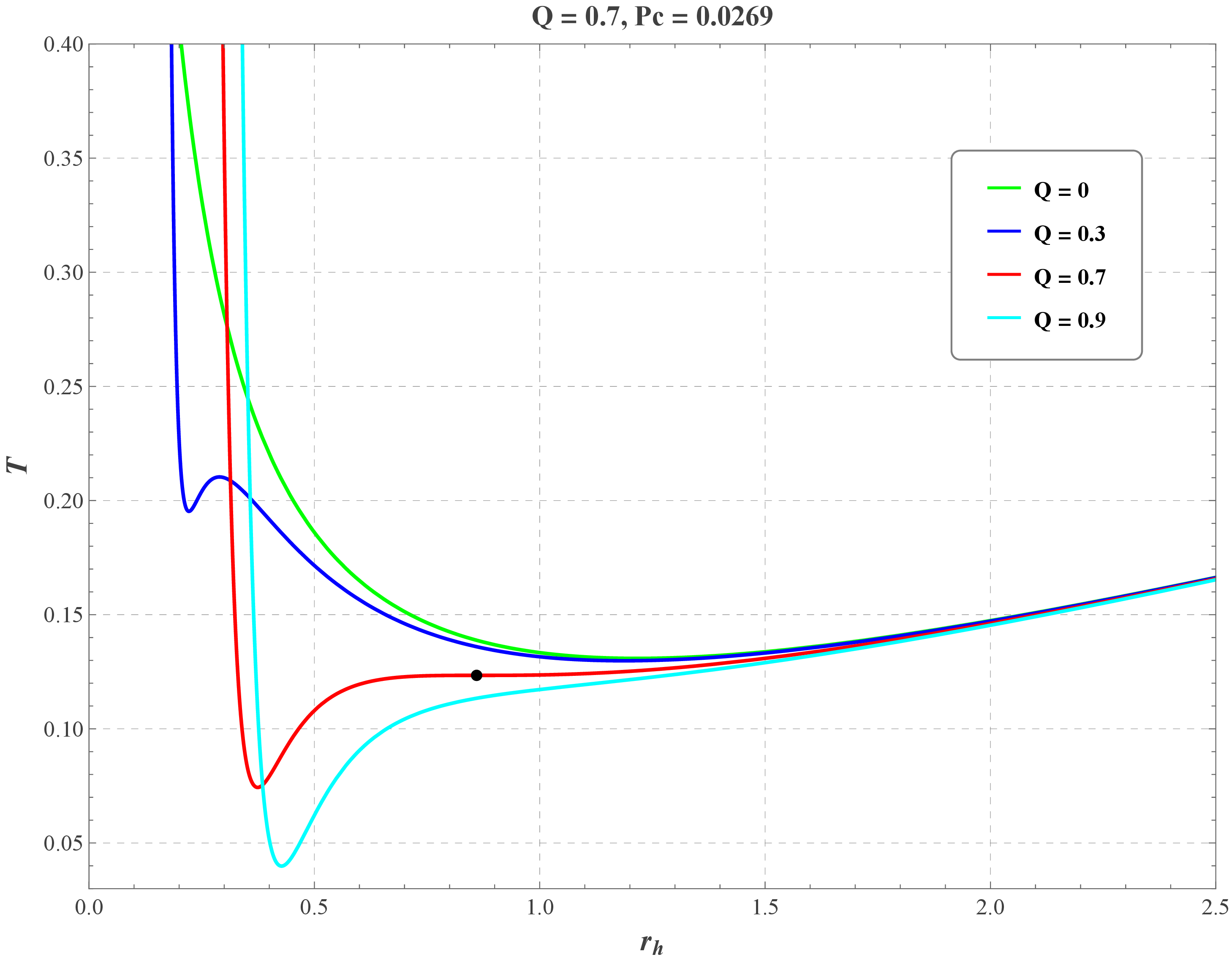}
    \caption{(a) Temperature as a function of horizon radius for fixed $Q = 0.7$ at pressures below, at, and above the critical pressure $P_c = 0.0269$. Below $P_c$, a van der Waals-type oscillation appears in the intermediate region, while the asymptotic behavior at small and large $r_h$ follows the Hawking--Page pattern with $T \to \infty$. (b) Temperature profiles for different values of the electric charge $Q$ at fixed subcritical pressure, illustrating how increasing $Q$ shifts the oscillatory feature to larger horizon radii and lowers the temperature minimum. There is a minimum temperature below which an AdS space exists.}
    \label{fig:temperature}
\end{figure}

The temperature behavior as a function of the horizon radius is shown in Fig.~\ref{fig:temperature}. The global structure resembles that of a Hawking--Page system: there is a single global minimum, below which only thermal AdS exists, and $T \to \infty$ in both limits $r_h \to 0$ and $r_h \to \infty$. Superimposed on this Hawking--Page background, a van der Waals-type oscillation appears in the intermediate region, reflecting the hybrid nature of the system.

The left panel shows $T$ vs. $r_h$ for fixed $Q = 0.7$ at pressures below, at, and above the critical pressure $P_c = 0.0269$ (see Table~\ref{tab:critical_points_selected}). Below $P_c$, an oscillatory branch emerges, signaling a first-order phase transition. At $P = P_c$, the oscillation degenerates into an inflection point, and above $P_c$, the temperature becomes monotonic. The right panel illustrates the $Q$-dependence at fixed subcritical pressure: increasing $Q$ shifts the transition to larger horizon radii and lowers the temperature minimum. For any nonzero $Q$, the temperature exhibits a single global minimum $T_{\min}$, with $T \to \infty$ as $r_h \to 0$ or $r_h \to \infty$; below $T_{\min}$, only thermal AdS exists. This hybrid structure, with a van der Waals-type oscillation superimposed on the global Hawking--Page skeleton, is a distinctive feature of the nonminimally coupled system.

\begin{figure}[H]
    \centering
    \subfloat[\(Q=0.7, P=P_c\)\label{fig:ft1}]{\includegraphics[width=0.30\textwidth]{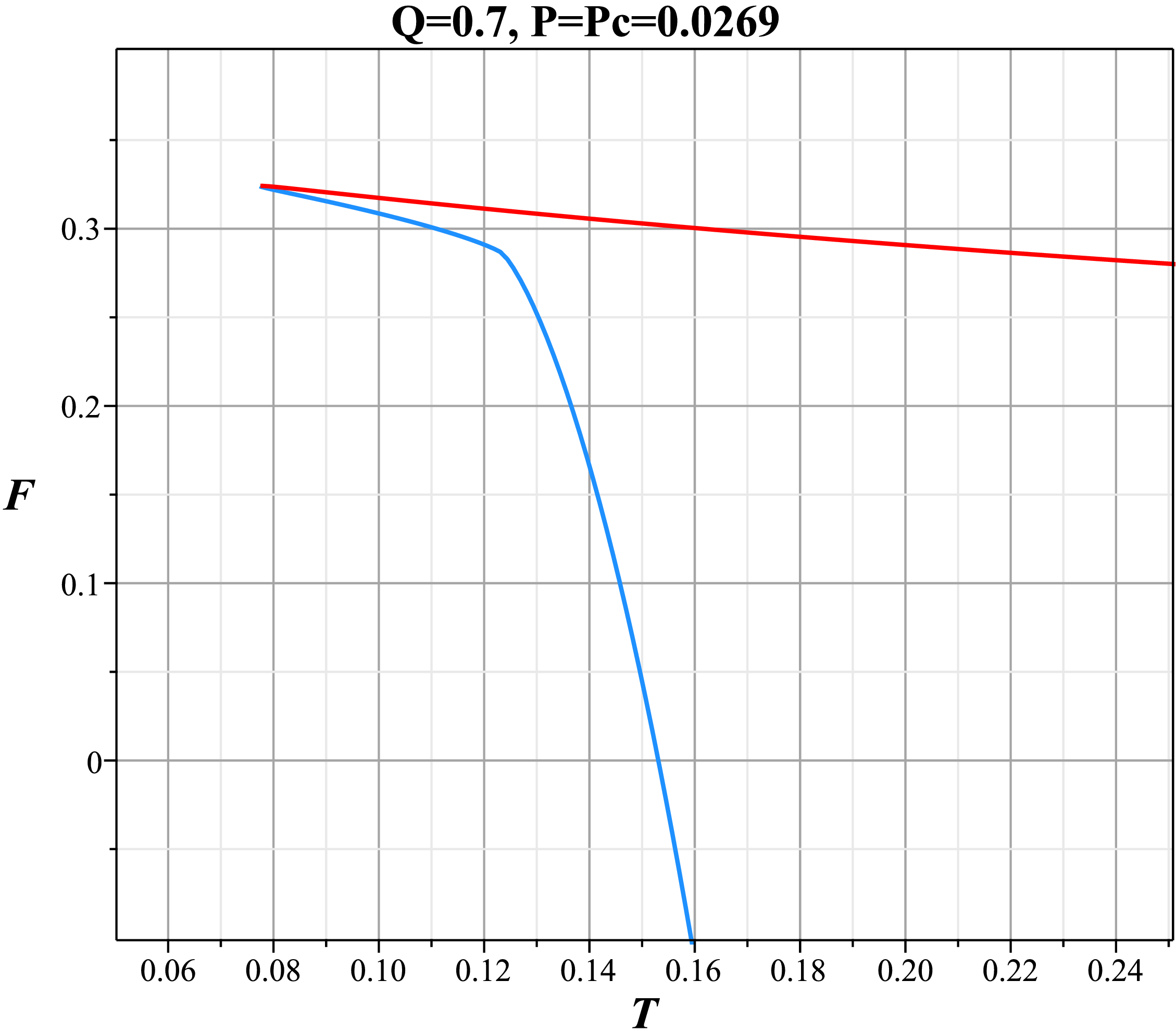}}
    \quad
    \subfloat[\(Q=0.7, P<P_c\)\label{fig:ft2}]{\includegraphics[width=0.30\textwidth]{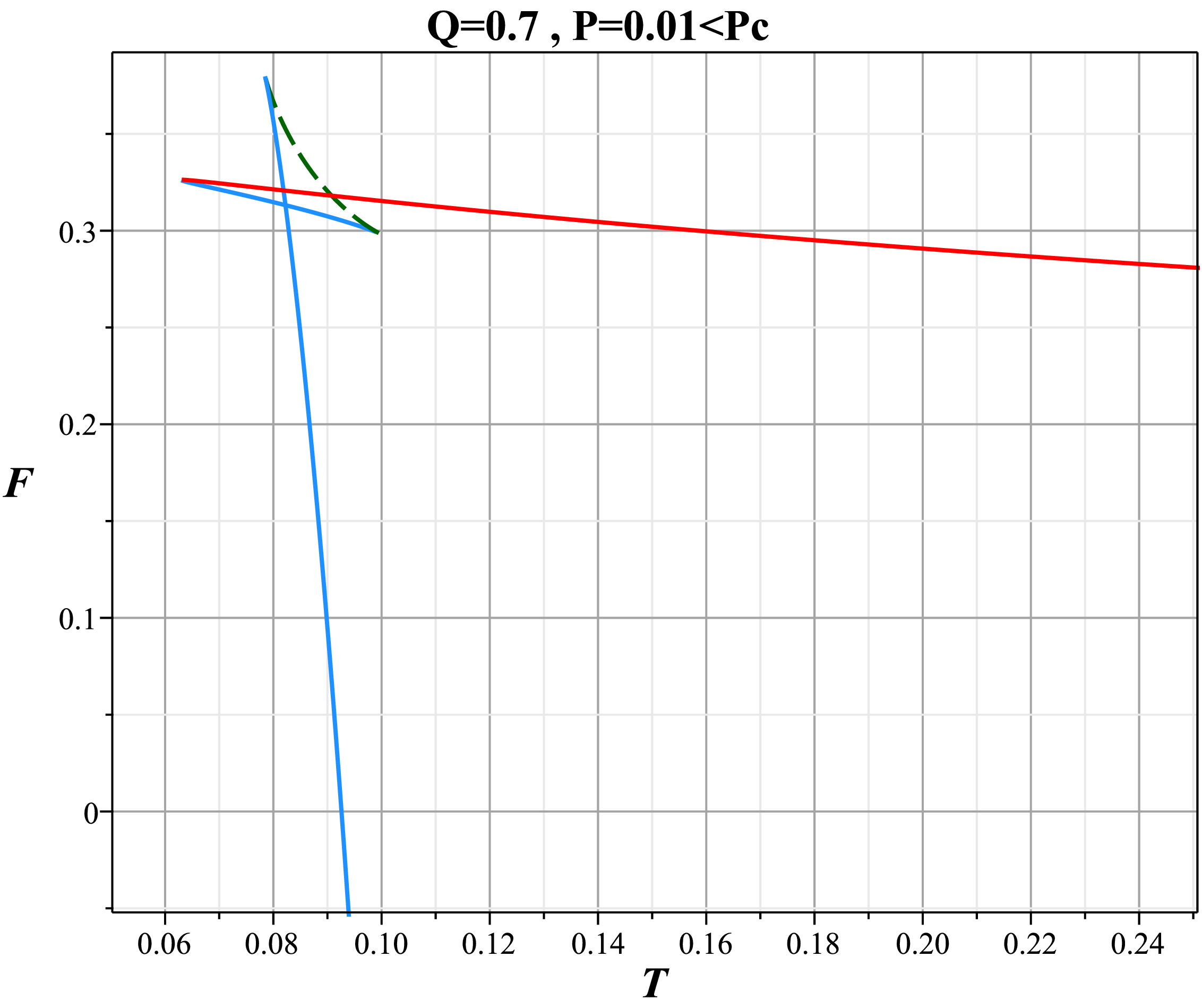}}
    \quad
    \subfloat[\(Q=0.7, P>P_c\)\label{fig:ft3}]{\includegraphics[width=0.30\textwidth]{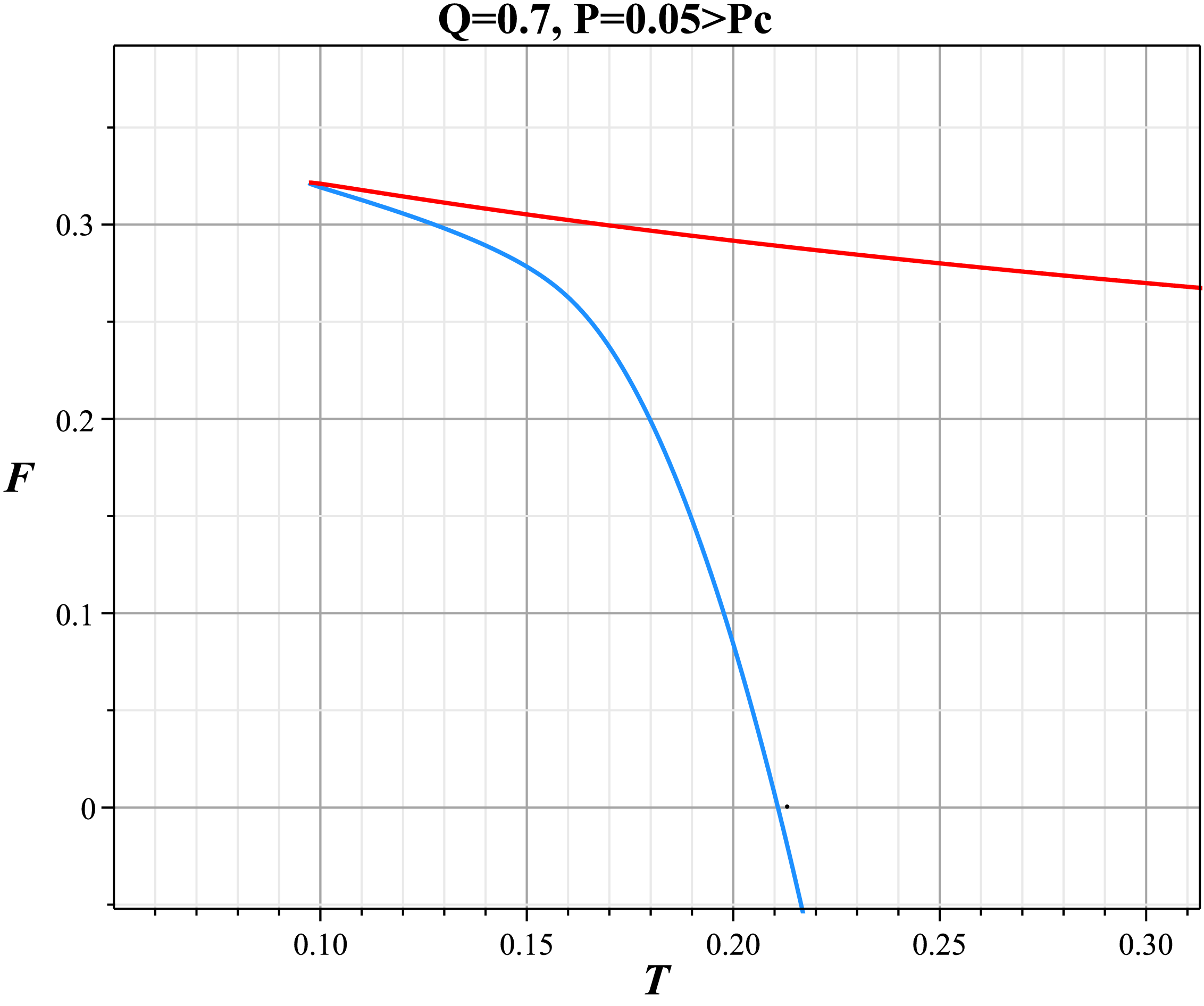}}
    
    \vspace{0.3cm}
    
    \subfloat[\(Q=1, P=P_c\)\label{fig:ft4}]{\includegraphics[width=0.30\textwidth]{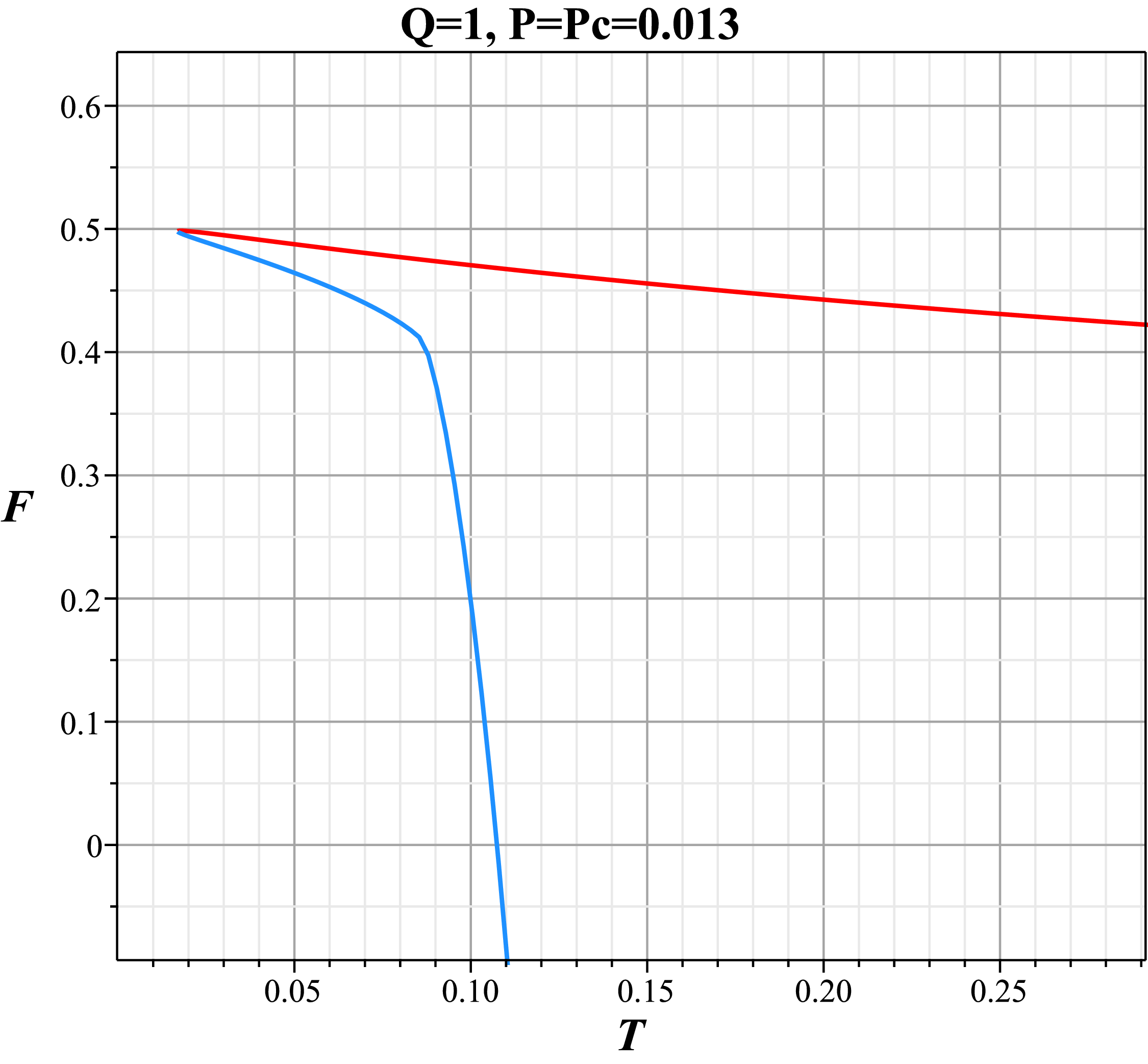}}
    \quad
    \subfloat[\(Q=1, P<P_c\)\label{fig:ft5}]{\includegraphics[width=0.30\textwidth]{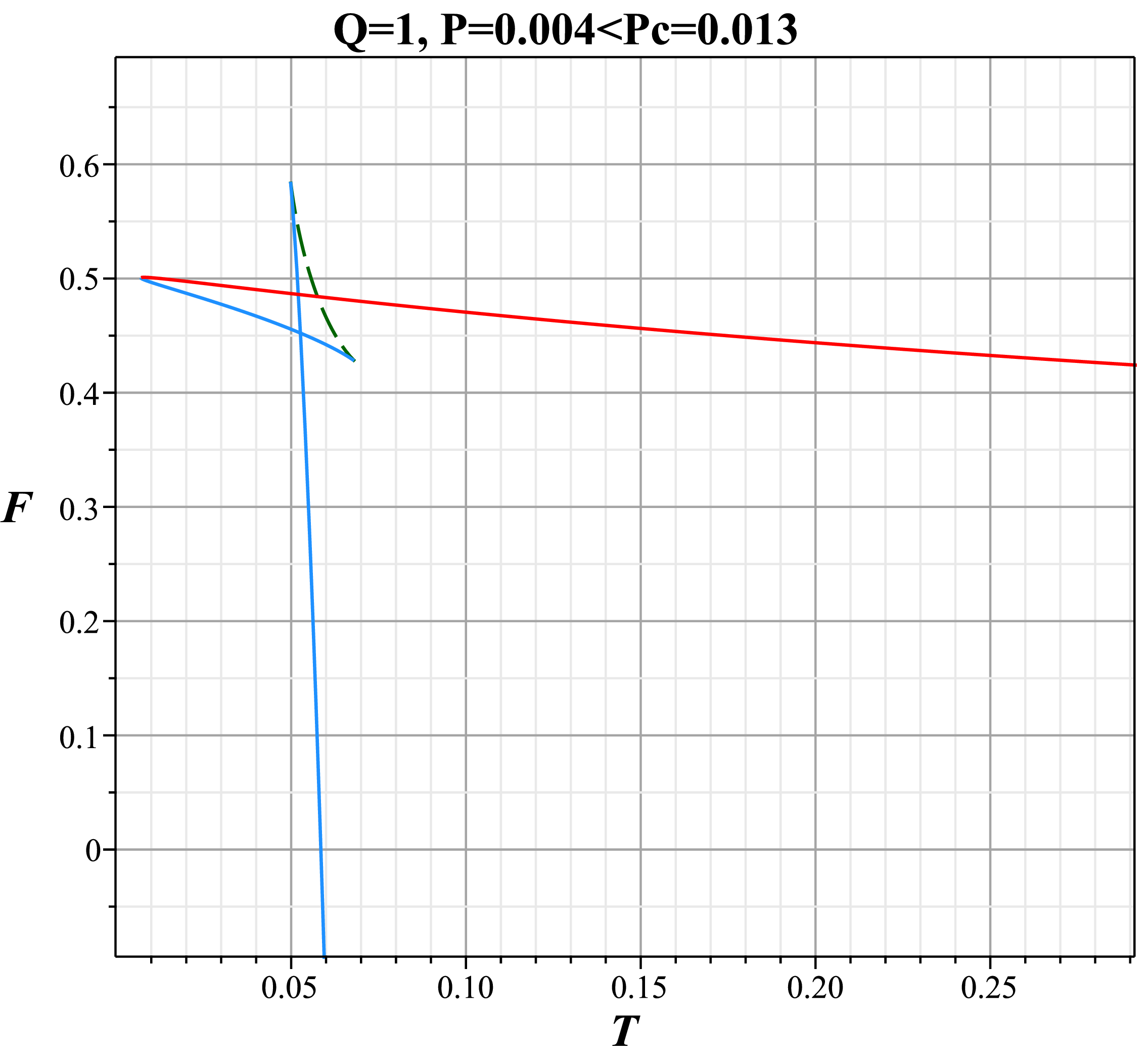}}
    \quad
    \subfloat[\(Q=1, P>P_c\)\label{fig:ft6}]{\includegraphics[width=0.30\textwidth]{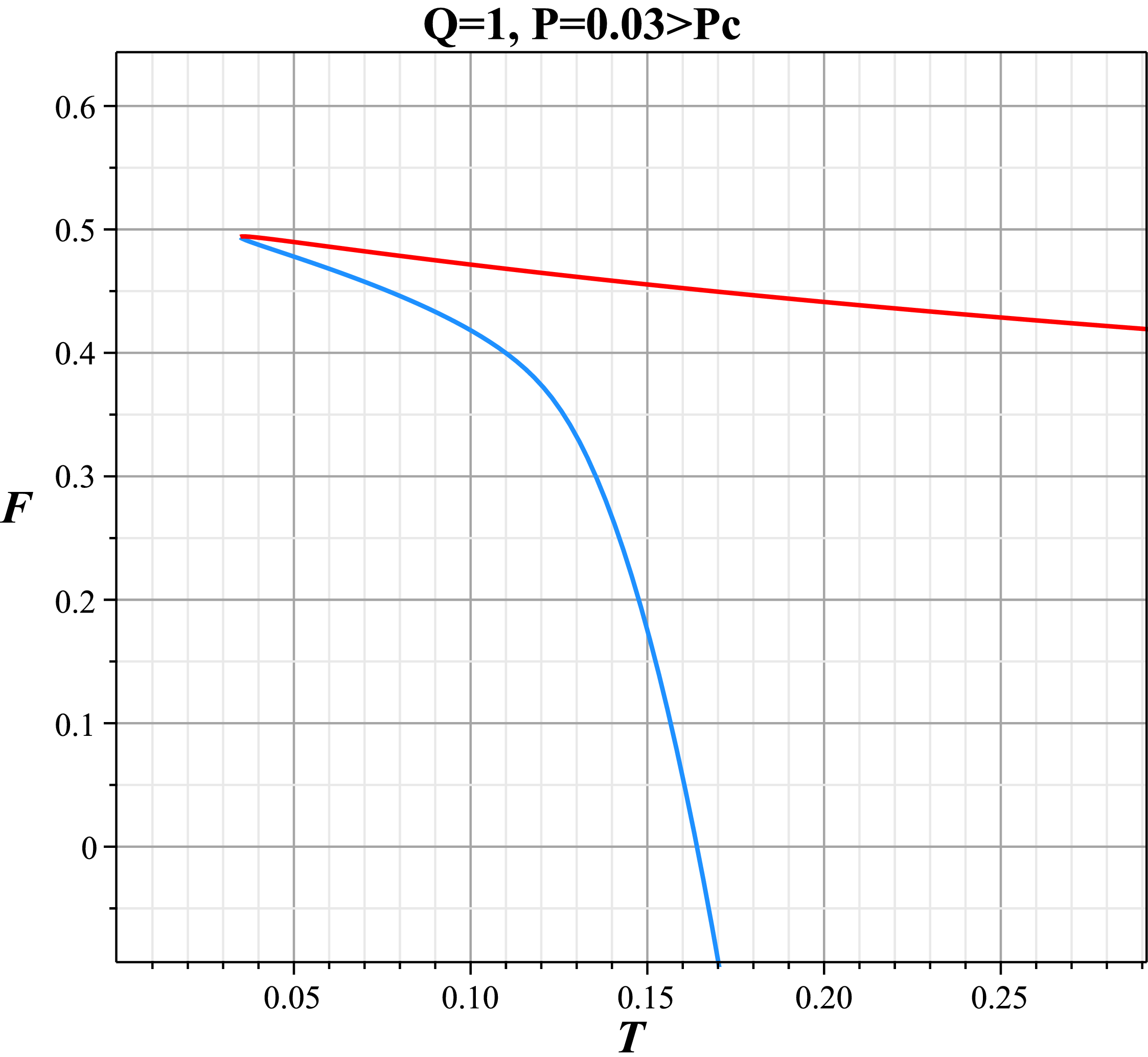}}
    
    \vspace{0.3cm}
    
    \subfloat[\(Q=0.4, P=P_c\)\label{fig:ft7}]{\includegraphics[width=0.30\textwidth]{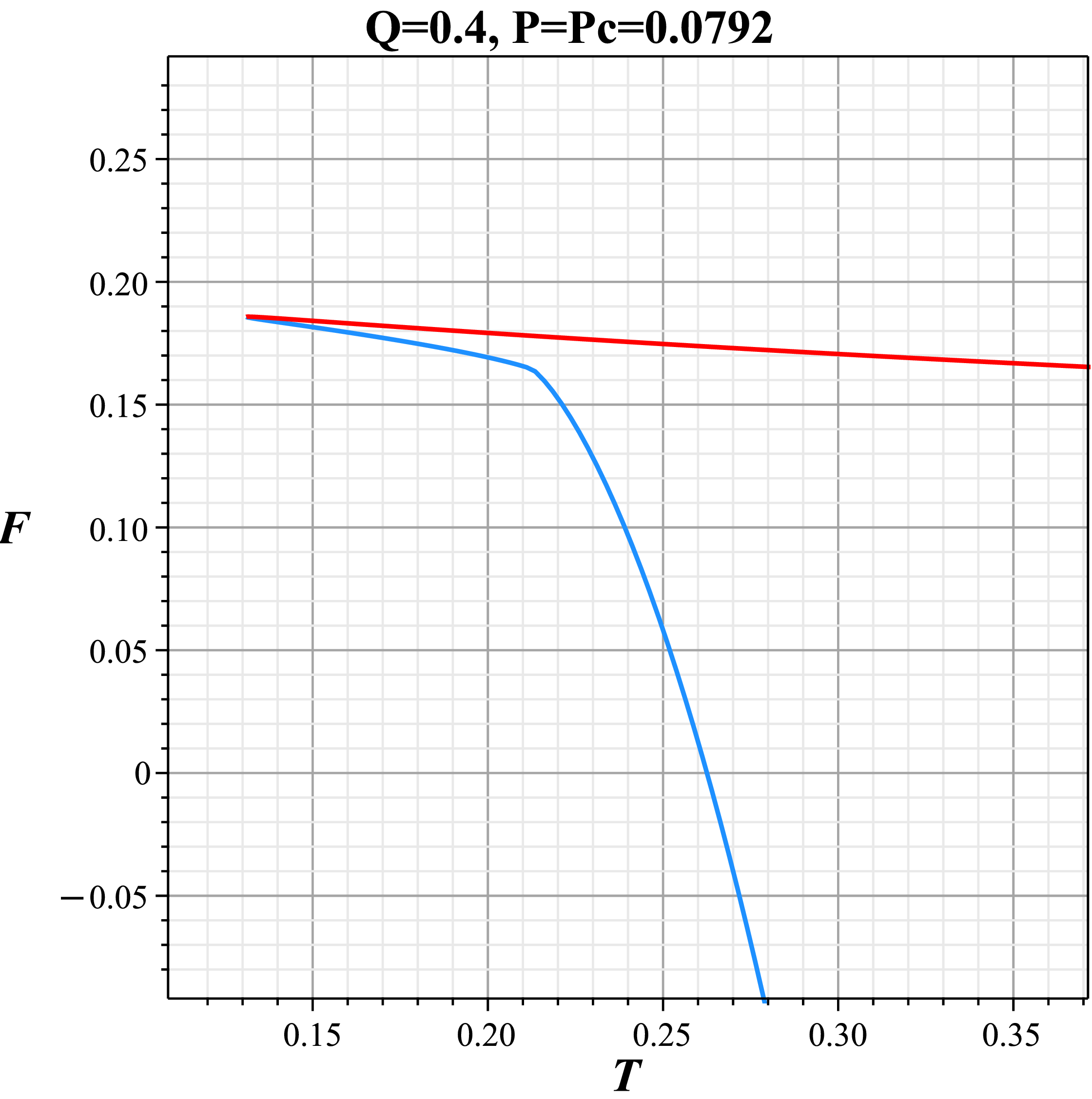}}
    \qquad
    \subfloat[\(Q=0.4, P<P_c\)\label{fig:ft8}]{\includegraphics[width=0.30\textwidth]{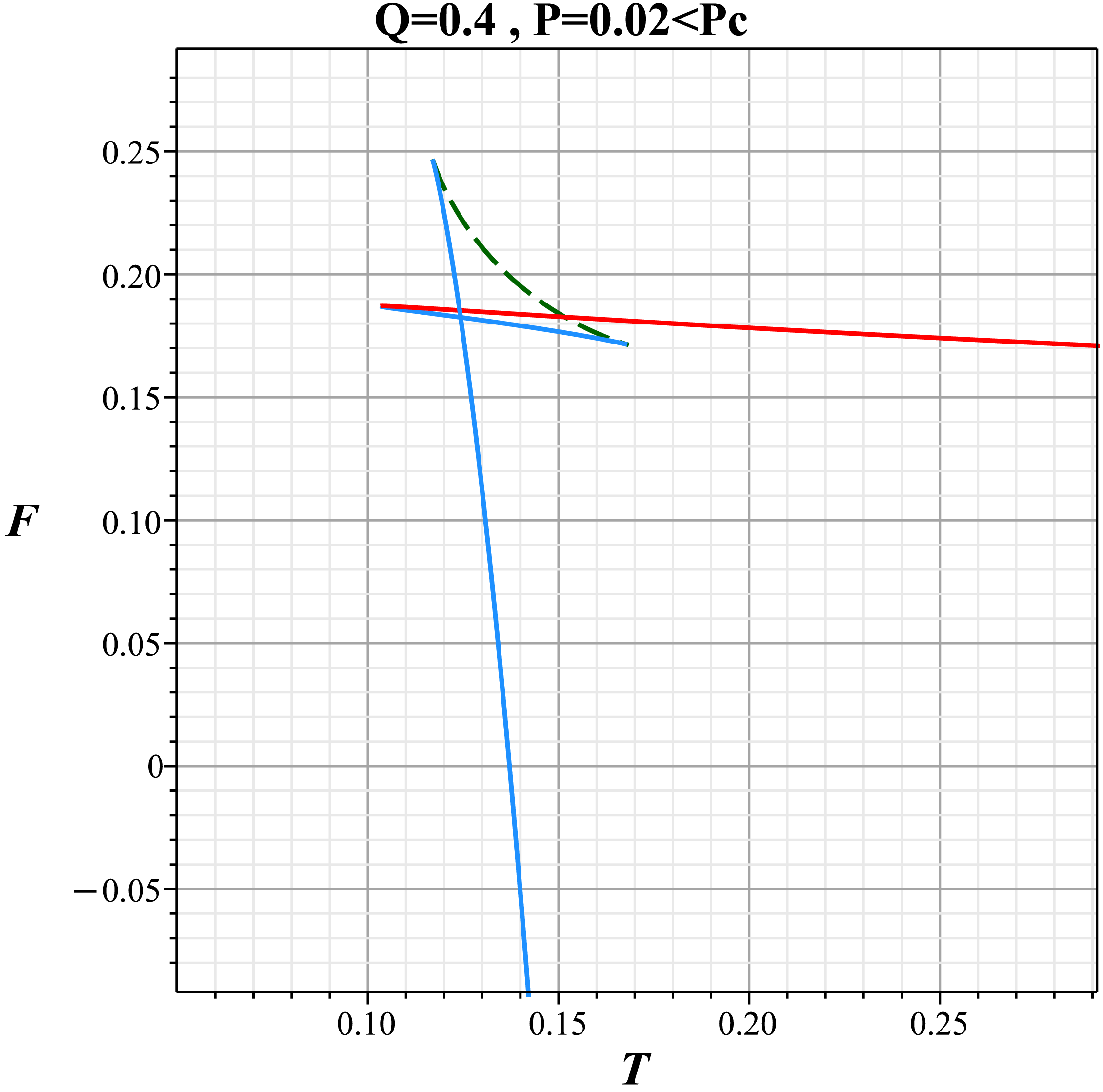}}
    \qquad
    \subfloat[\(Q=0.4, P>P_c\)\label{fig:ft9}]{\includegraphics[width=0.30\textwidth]{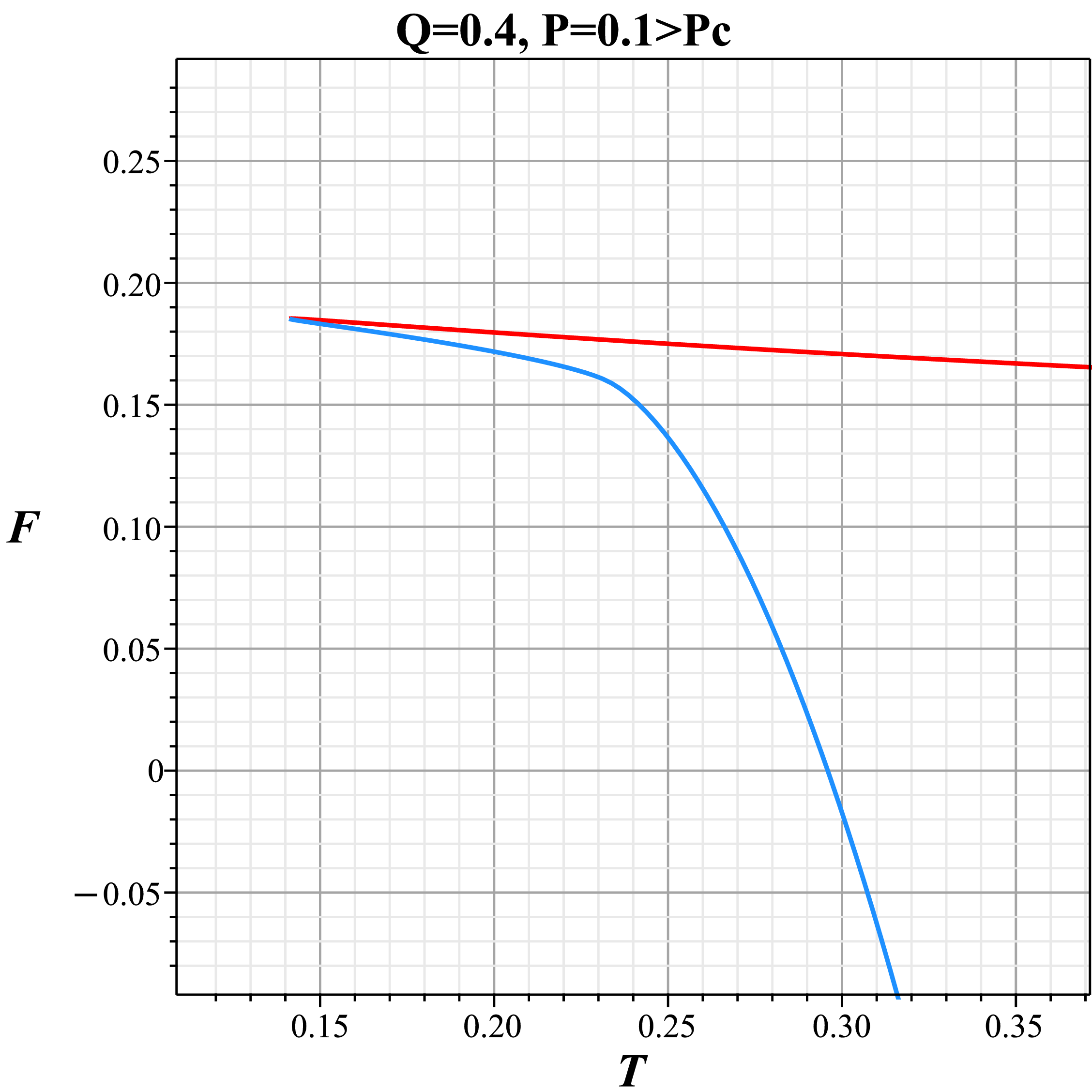}}
    
    \caption{Helmholtz free energy $F$ as a function of temperature $T$ for different values of $Q$ and $P$. Rows correspond to (top) $Q=0.7$, (middle) $Q=1$, and (bottom) $Q=0.4$. Columns correspond to (left) $P=P_c$, (middle) $P<P_c$, and (right) $P>P_c$. The swallowtail structure in the middle column indicates a first-order phase transition.}
    \label{fig:FT}
\end{figure}

The Helmholtz free energy further confirms this intriguing pattern, as shown in Fig.~\ref{fig:FT}. The free energy exhibits a global Hawking--Page structure consisting of an unstable upper branch and a stable lower branch. Remarkably, a van der Waals-type oscillation in the form of a swallowtail appears on the lower branch, embedding a local phase transition within the global HP framework.

The left panels show $F(T)$ at the critical pressure. Two branches are present: an upper unstable branch and a lower stable branch, with a characteristic kink appearing on the lower branch at $P = P_c$. The middle panels correspond to pressures below $P_c$, where a swallowtail emerges on the lower stable branch, signaling a first-order phase transition between small and intermediate black holes. This transition occurs entirely within the globally stable HP branch. The right panels depict the case $P > P_c$, where the free energy becomes smooth and no swallowtail is present, indicating the absence of a first-order transition.

This hybrid structure is further illuminated by the following interpretation. There is an unstable upper branch and a stable lower branch, crossing at the HP temperature. On the lower stable branch, a swallowtail appears when $P < P_c$, signaling a small-to-intermediate black hole first-order transition that occurs entirely within the HP stable phase. As the black hole evaporates (i.e., as the temperature increases), it first undergoes a VdW-type phase transition (small $\to$ intermediate) while remaining globally stable; only at higher temperatures does it jump to the large black hole branch. The system thus exhibits two distinct phase transitions in different temperature ranges, both controlled by the same coupling $\epsilon$.

The local stability of the system can be examined via its heat capacity. The heat capacity at constant pressure is:
\begin{equation}\label{CC}
C_P = \left(\left( \frac{d\mathcal{H}}{dr_h} \right) \left( \frac{dT}{dr_h} \right)^{-1}\right)_{P,Q}
= \frac{
  \left(
    -\frac{64 Q^2 \pi^2 P^2}{r_h^2}
    - \frac{2 Q^4 \pi P}{r_h^6}
    + \frac{Q^6}{4 r_h^{10}}
    - \frac{12 Q^2 \pi P}{r_h^4}
    - \frac{7 Q^4}{8 r_h^8}
  \right) \epsilon
  + 4 r_h^2 \pi P
  - \frac{Q^2}{8 r_h^2}
  + \frac{1}{2}
}{
  2P
  + \frac{3 Q^2}{16 \pi r_h^4}
  - \frac{1}{4 \pi r_h^2}
  + \left(
    -\frac{7 Q^4 P}{r_h^8}
    - \frac{11 Q^6}{32 \pi r_h^{12}}
    + \frac{10 Q^2 P}{r_h^6}
    + \frac{9 Q^4}{16 \pi r_h^{10}}
  \right) \epsilon
}
\end{equation}

\begin{figure}[H]
    \centering
    \includegraphics[width=0.31\textwidth]{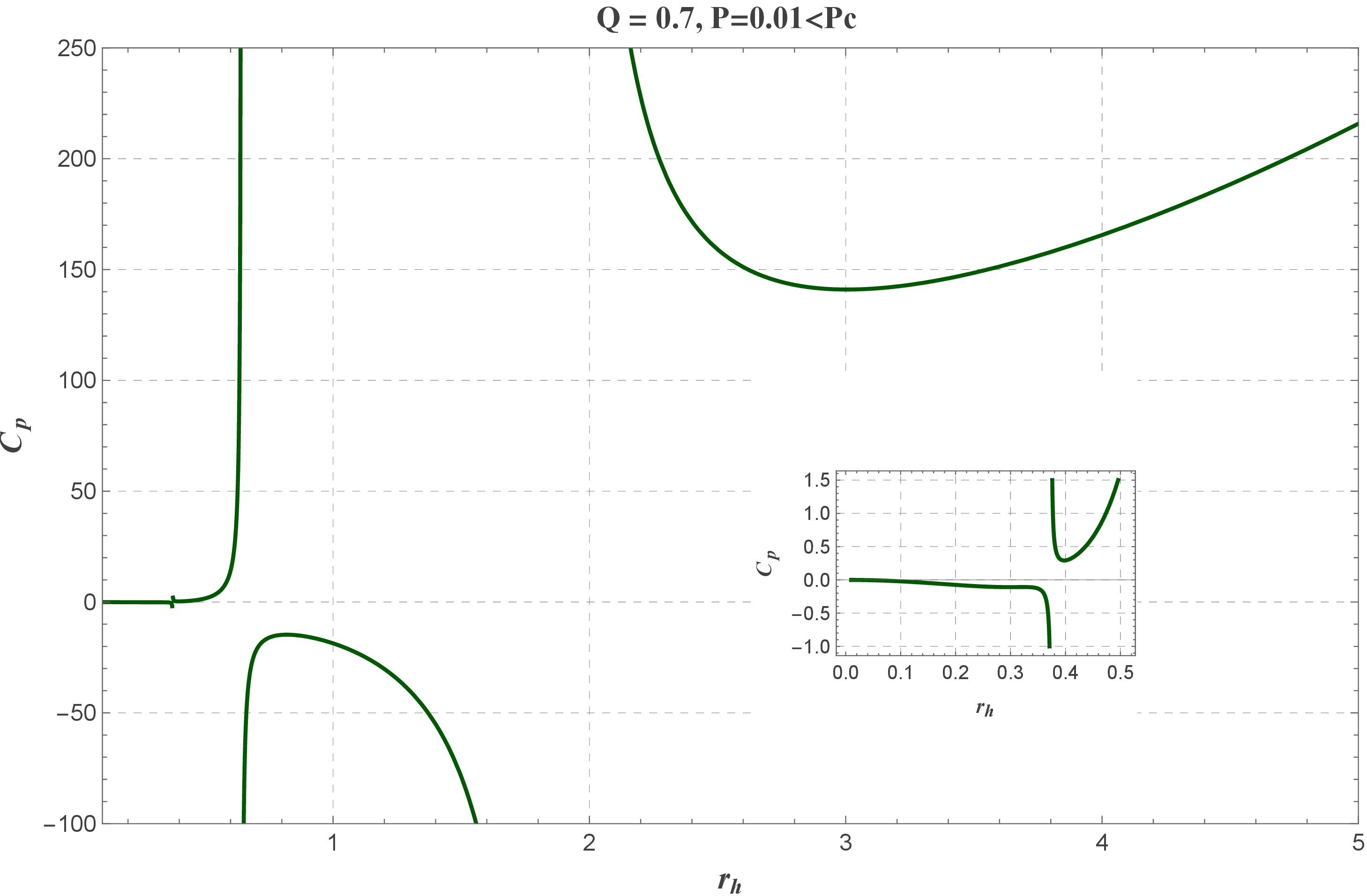}
    \quad
    \includegraphics[width=0.31\textwidth]{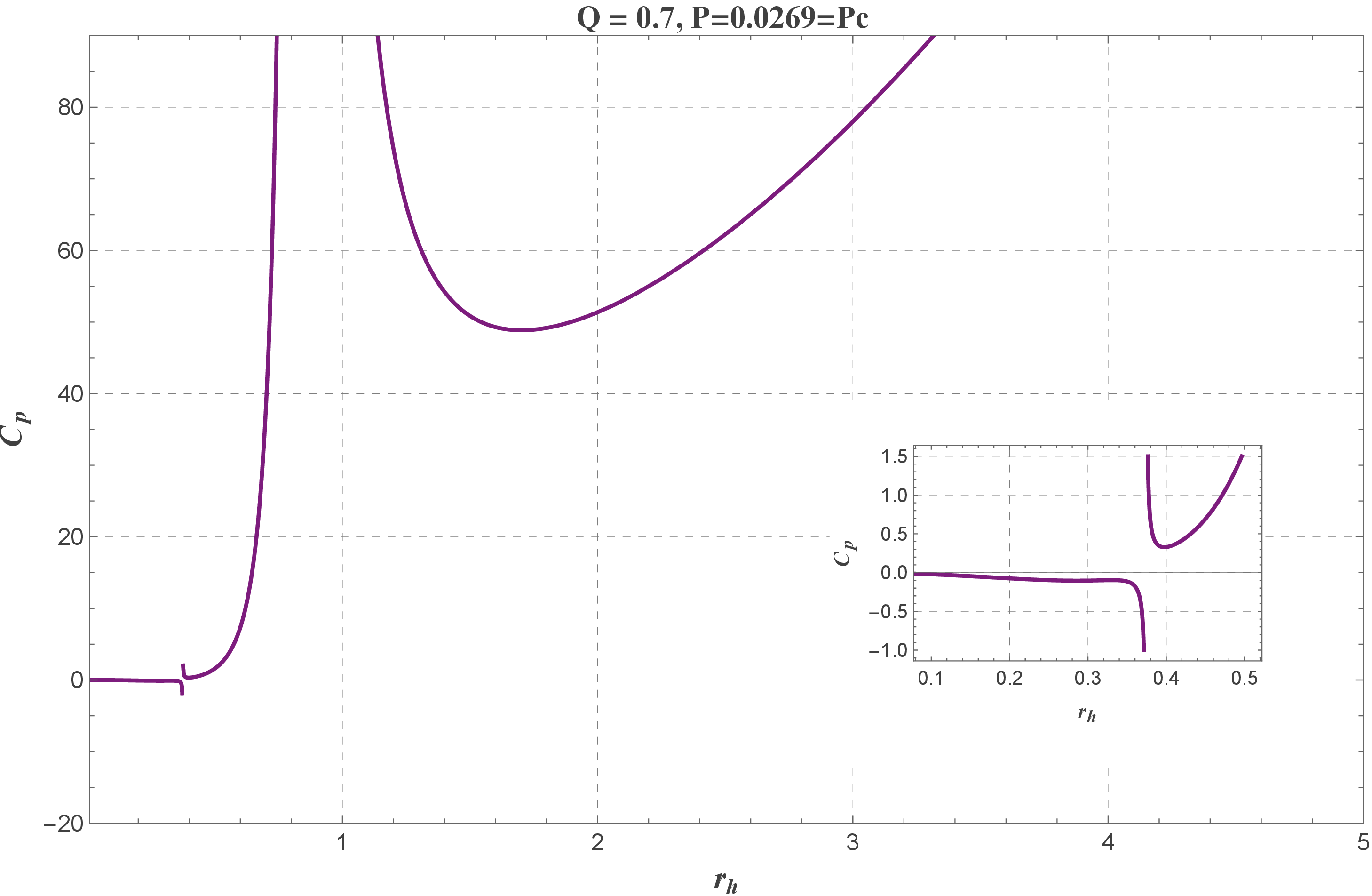}
    \quad
    \includegraphics[width=0.31\textwidth]{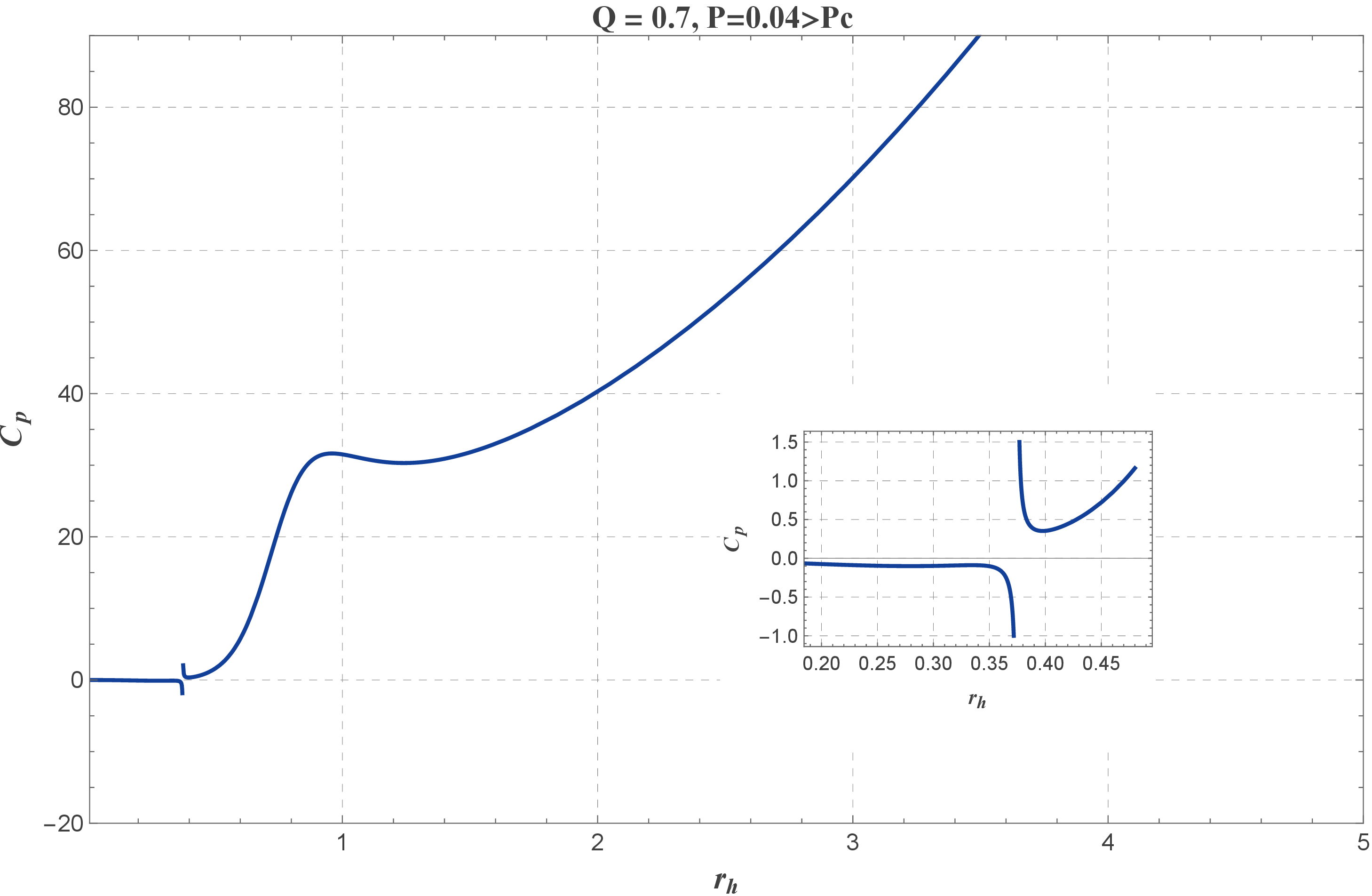}
    \caption{(a) Heat capacity $C_P$ as a function of horizon radius $r_h$ for $Q = 0.7$ and $P = 0.01 < P_c = 0.0269$. Three divergences separate four regions with alternating stability: unstable, stable, unstable, stable, reflecting the hybrid Hawking--Page and van der Waals nature. The first divergence at small $r_h$ is characteristic of the HP transition. (b) Heat capacity at the critical pressure $P = P_c$, where the two intermediate divergences merge into a single critical point. (c) Heat capacity for $P > P_c$, where the intermediate unstable region vanishes, leaving only the HP-induced divergence at small horizons. Insets magnify the small-horizon region for clarity.}
    \label{fig:capacity1}
\end{figure}

Since the present system exhibits both global Hawking--Page behavior and local van der Waals behavior, signatures of both phenomena are expected in the heat capacity, as illustrated in Fig.~\ref{fig:capacity1}.

The left panel shows $C_P$ for $Q = 0.7$ and $P = 0.01 < P_c = 0.0269$. Four distinct regions emerge, characterized by an alternating stability pattern: unstable ($C_P < 0$), stable ($C_P > 0$), unstable, and stable, separated by three divergences. The first divergence, occurring at small horizon radii, is a hallmark of the Hawking--Page structure. This alternating sequence confirms the presence of the global HP skeleton with embedded VdW oscillations.

The middle panel depicts the heat capacity at the critical pressure $P = P_c$, where the two intermediate divergences coalesce, signaling the merging of the van der Waals transition into a critical point. The right panel corresponds to the above-critical case $P > P_c$. For clarity, the small-horizon divergences are highlighted in magnified insets within each panel.

This heat capacity structure reveals a richer stability pattern than either pure HP or standard VdW systems. Below $P_c$, one finds four distinct regions with alternating stability: ultra-small unstable, small stable, intermediate unstable, and large stable.

The thermodynamic analysis reveals that the nonminimal coupling $\epsilon F^{\alpha\beta}F^{\gamma\lambda}R_{\alpha\gamma}R_{\beta\lambda}$ produces a hybrid thermodynamic system. This system does not belong purely to the VdW type nor purely to the HP type. Instead, it embeds a VdW-like liquid--gas phase transition inside the global HP framework. The electric charge $Q$ acts as a knob that continuously tunes the strength of the VdW-type transition relative to the HP background. For $Q = 0$, the system reduces to pure HP -- no VdW loop, no swallowtail. As $Q$ increases, the VdW loop grows, and the critical pressure  decreases (Table~\ref{tab:critical_points_selected}). For fixed $\epsilon$, increasing $Q$ pushes the VdW transition to larger horizon radii and lower temperatures. The nonminimal coupling $\epsilon$ makes this tuning possible even at small $Q$. In standard Einstein--Maxwell AdS black holes, the HP transition occurs for neutral black holes while the VdW transition occurs for charged black holes, but the global topology of $T(r_h)$ changes (single minimum vs. multi-extrema). Here, thanks to $\epsilon$, the system retains the HP global shape while simultaneously exhibiting a VdW local transition. This is unusual and suggests that the nonminimal coupling stabilizes the HP skeleton, allowing the VdW physics to appear as a defect on the stable branch. Consequently, for all nonzero values of $Q$, the system remains a hybrid, with the HP structure always present. The coupling $\epsilon$ controls the coexistence of both phenomena, but the system never becomes a pure VdW system, regardless of how large $Q$ becomes.

\section{Topological formalism for phase transitions}
\label{sec4}

The central idea of the topological approach is to view each black hole solution as a topological defect in the thermodynamic parameter space. This is achieved by constructing a generalized off-shell free energy that depends on both the horizon radius $r_h$ and an auxiliary inverse temperature parameter $\tau$, which originates from the cavity parameter in the Hawking--York Euclidean path integral approach. The standard on-shell free energy is recovered when $\tau = \beta = 1/T$.

A two-component vector field $\phi = (\phi^{r_h}, \phi^\Theta)$ is defined as
\begin{equation}
\phi^{r_h} = \frac{\partial \mathcal{F}}{\partial r_h}, \qquad 
\phi^\Theta = -\frac{\cos \Theta}{\sin^2 \Theta},
\label{eq:vector_field}
\end{equation}
where $\mathcal{F}(r_h, \tau) = M(r_h) - S(r_h)/\tau$ is the generalized free energy for a black hole of mass $M$ (enthalpy in extended phase space) and entropy $S$ enclosed in a cavity at fixed temperature $1/\tau$, and $\Theta \in (0,\pi)$ is an auxiliary angular coordinate introduced to apply Duan's topological current $\phi$-mapping theory \cite{Duan1984, Duan1979}. 

The zero points of $\phi^{r_h}$ (i.e., $\partial \mathcal{F}/\partial r_h = 0$) correspond to the black hole solutions. 

\subsection{Winding number and topological charge}

Following Duan's $\phi$-mapping topological current theory, we introduce the unit vector field
\begin{equation}
n^a=\frac{\phi^a}{\|\phi\|},
\end{equation}
where $a=(r_h,\Theta)$. The conserved topological current is defined by
\begin{equation}
j^\mu=
\frac{1}{2\pi}
\epsilon^{\mu\nu}\epsilon_{ab}
\partial_\nu n^a\partial_\rho n^b,
\qquad
\mu,\nu,\rho=0,1,
\label{eq:topocurrent}
\end{equation}
which satisfies $\partial_\mu j^\mu=0$. This current can be expressed as
\begin{equation}
j^\mu=
\delta^2(\phi)
J^\mu\left(\frac{\phi}{x}\right),
\label{eq:topocurrent2}
\end{equation}
showing that it is localized at the zero points of $\phi$, where $J^\mu(\phi/x)$ denotes the Jacobian vector.

Consequently, the winding number associated with the $i$th zero point is
\begin{equation}\label{wi}
w_i=\int j^0\,d^2x=\beta_i\eta_i,
\end{equation}
where
\begin{equation}
j^0=
\beta_i\eta_i\,
\delta^2(\vec{x}-\vec{z}_i).
\end{equation}
Here, the positive integer $\beta_i$ is the Hopf index, which counts the number of times the mapping $\phi^a$ wraps around its internal space as $x^\mu$ encircles the zero point $z_i$, while
\begin{equation}
\eta_i=
\mathrm{sign}
\left[
J^0\left(\frac{\phi}{x}\right)_{z_i}
\right]
=\pm1
\end{equation}
is the Brouwer degree, determining the orientation of the vector field around the zero.

For isolated nondegenerate zero points, the Hopf index satisfies $\beta_i=1$, and therefore
\begin{equation}\label{nd}
w_i=\eta_i=
\mathrm{sign}
\left(
\frac{\partial^2\mathcal F}
{\partial r_h^2}
\right)
\Bigg|_{r_h=\mathrm{ZP}_i}.
\end{equation}
Here, "nondegenerate" means that the zero points of the vector field are isolated and the Jacobian determinant at each zero point is nonzero, ensuring that the winding number is well-defined and takes only the values $\pm 1$. Hence, a positive (negative) second derivative corresponds to a winding number $w_i=+1$ ($w_i=-1$), identifying a locally stable (unstable) black-hole branch.

For numerical calculations, the winding number may equivalently be obtained from the deflection angle of the normalized vector field along a closed contour $C$ enclosing a given zero point,
\begin{equation}\label{winding}
w=
\frac{1}{2\pi}
\oint_C
\epsilon_{ab}
n^a\,dn^b
=
\frac{1}{2\pi}
\int_0^{2\pi}
\epsilon_{ab}
n^a
\frac{\partial n^b}{\partial\vartheta}\,
d\vartheta,
\end{equation}
where $\vartheta$ parametrizes the contour. In practice, we choose the contour to be traversed counterclockwise by convention, with the parametrization
\begin{equation}
\label{eq:contour_param}
\begin{aligned}
r_h(\vartheta)&=a\cos\vartheta+z_0,\\
\Theta(\vartheta)&=b\sin\vartheta+\frac{\pi}{2},
\end{aligned}
\qquad
0\leq\vartheta\leq2\pi,
\end{equation}
where the parameters $a$ and $b$ are selected such that the contour encloses the desired zero point $z_0$.

The global topological number is obtained by summing over all isolated zero points,
\begin{equation}
W=\sum_{i=1}^{N}w_i,
\label{eq:totalwinding}
\end{equation}
where $N$ denotes the total number of zeros in the region of interest. Since $W$ is invariant under continuous deformations of the thermodynamic potential, it provides a robust topological classification of black-hole phases.\cite{Wei:2021vdx}

\subsection{Known topological classes}

Based on the asymptotic behavior of the inverse temperature
$\beta(r_h)=1/T(r_h)$ near the minimal horizon radius $r_m$ and in the limit
$r_h\rightarrow\infty$, black-hole systems can be classified into four universal classes \cite{Wei:2024gfz},
\[
W^{1-},\quad W^{0+},\quad W^{0-},\quad W^{1+},
\]
characterized by the following asymptotic behaviors:

\begin{itemize}
    \item $W^{1-}$: $\beta(r_m)=0$, $\beta(\infty)=\infty$ (e.g., Schwarzschild black holes).
    \item $W^{0+}$: $\beta(r_m)=\infty$, $\beta(\infty)=\infty$ (e.g., Reissner--Nordstr\"om black holes).
    \item $W^{0-}$: $\beta(r_m)=0$, $\beta(\infty)=0$ (e.g., Schwarzschild--AdS black holes).
    \item $W^{1+}$: $\beta(r_m)=\infty$, $\beta(\infty)=0$ (e.g., Reissner--Nordstr\"om--AdS black holes).
\end{itemize}

In this classification scheme, the stability of the innermost branch determines the low-temperature behavior, whereas the outermost branch governs the high-temperature limit. Accordingly, the global topological number $W$ may take the values $0$, $+1$, or $-1$.

Complementary to the asymptotic behavior of $\beta(r_h)$, the topological class can also be inferred from the direction of the unit vector field $\phi$ on the boundaries of the parameter space. To systematically analyze this, we consider the closed boundary contour $C = I_1 \cup I_2 \cup I_3 \cup I_4$, which encompasses the entire parameter region $(r_h, \Theta)$, where $r_h \in (r_m, \infty)$ and $\Theta \in (0, \pi)$. The four segments are defined as follows:
\begin{align}
I_1 &= \{r_h = \infty,\; \Theta \in (0, \pi)\}, \nonumber\\
I_2 &= \{r_h \in (\infty, r_m),\; \Theta = \pi\}, \nonumber\\
I_3 &= \{r_h = r_m,\; \Theta \in (\pi, 0)\}, \nonumber\\
I_4 &= \{r_h \in (r_m, \infty),\; \Theta = 0\}. \nonumber
\end{align}
On the segments $I_2$ and $I_4$, the vector field $\phi$ is orthogonal to the boundary by construction, so the asymptotic behavior of interest is along $I_1$ and $I_3$ \cite{Wei:2024gfz}. On these segments, the $r_h$-component of $\phi$ is given by $\phi^{r_h}=(\frac{\partial S}{\partial r_h}) (1/\beta - 1/\tau)$, where $\tau$ is the fixed cavity temperature and we have assumed $\frac{\partial S}{\partial r_h}>0$. Thus, as $\beta \to 0$, the vector field points rightward ($\phi^{r_h}>0$), while as $\beta \to \infty$, it points leftward ($\phi^{r_h}<0$). The directions of $\phi$ on all four boundary segments for each topological class are summarized in Table~\ref{tab:topo_boundary} below.

\begin{table}[H]
\centering
\caption{Direction of unit vector field  on the boundary segments $I_1, I_2, I_3, I_4$ for each topological class, along with the global topological number $W$ and representative examples.}
\label{tab:topo_boundary}
\begin{tabular}{lcccccc}
\toprule
Class & $I_1$ & $I_2$ & $I_3$ & $I_4$ & $W$ & Example \\
\midrule
$W^{1-}$ & $\leftarrow$ & $\uparrow$ & $\rightarrow$ & $\downarrow$ & $-1$ & Schwarzschild \\
$W^{0+}$ & $\leftarrow$ & $\uparrow$ & $\leftarrow$ & $\downarrow$ & $0$ & RN (fixed $Q$) \\
$W^{1+}$ & $\rightarrow$ & $\uparrow$ & $\leftarrow$ & $\downarrow$ & $+1$ & RN-AdS \\
$W^{0-}$ & $\rightarrow$ & $\uparrow$ & $\rightarrow$ & $\downarrow$ & $0$ & Schwarzschild-AdS \\
\bottomrule
\end{tabular}
\end{table}

This boundary-direction analysis provides a direct visual criterion for identifying the topological class of a given black hole system, complementing the $\beta(r_h)$ asymptotics \cite{WeiLiuMann2022,Wei:2024gfz}.

\subsection{Extensions to new classes and subclasses}

Subsequent studies \cite{Wu:2022whe, ZhuWu2024, WuGu2024, Chen:2024sow} revealed that several black-hole solutions, including static charged AdS black holes in gauged supergravity, dyonic AdS black holes, and multi-charge AdS black holes, do not fit into the original fourfold classification. These developments uncovered a richer topological structure beyond the known classes, including

\begin{itemize}
    \item a new class $W^{0\leftrightarrow1+}$, in which the global topological number changes from $W=0$ to $W=+1$ with increasing temperature, signaling a temperature-driven topological phase transition;

    \item two additional subclasses, $\widetilde{W}^{1+}$ and $\widehat{W}^{1+}$, both characterized by $W=+1$ but distinguished by different sequences of stable and unstable branches and distinct low-temperature behaviors.
\end{itemize}

Table~\ref{tab:topo_classes} summarizes both the original and the extended topological classifications.

\begin{table}[H]
\centering
\caption{Summary of topological classes in black-hole thermodynamics. Here, $W$ denotes the global topological number, $r_m$ is the minimal horizon radius, and $\beta=1/T$.}
\label{tab:topo_classes}
\begin{tabular}{lcccc}
\toprule
Class & $W$ & $\beta(r_m)$ & $\beta(\infty)$ & Example \\
\midrule
$W^{1-}$ & $+1$ & $0$ & $\infty$ & Schwarzschild \\
$W^{0+}$ & $0$ & $\infty$ & $\infty$ & RN (fixed $Q$) \\
$W^{0-}$ & $0$ & $0$ & $0$ & Schwarzschild-AdS \\
$W^{1+}$ & $+1$ & $\infty$ & $0$ & RN-AdS \\
\midrule
$W^{0\leftrightarrow1+}$ & $0$ or $+1$ & finite & $0$ & Static two-charge AdS (gauged SUGRA) \\
$\widetilde{W}^{1+}$ & $+1$ & finite & $0$ & EMDA-AdS (large $Q$) \\
$\widehat{W}^{1+}$ & $+1$ & $\infty$ & $0$ & Dyonic AdS \\
\bottomrule
\end{tabular}
\end{table}

This topological framework has subsequently been applied to a variety of black-hole systems, including rotating solutions \cite{Wu:2022whe}, Gauss--Bonnet gravity \cite{LiuWang2023} and Born--Infeld electrodynamics \cite{Ali2024}. Moreover, it gives rise to a bulk-boundary correspondence, whereby the topology of the dual CFT thermodynamics coincides with that of the bulk AdS black hole in the canonical ensemble \cite{Ali2024,Zhang:2023uay}.

The topological class may change as the charge is varied. For a nonminimal gauge-curvature coupling, as studied in Ref.~\cite{Rahmani:2025iks}, it has been shown that above a certain value of the charge, the system belongs to the $W^{1+}$ class, while below it, it belongs to the $W^{0-}$ class.

We now apply this framework to the nonminimally coupled $F^{\alpha\beta}F^{\gamma\lambda}R_{\alpha\gamma}R_{\beta\lambda}$ AdS black hole. In particular, we show that the coupling parameter $\epsilon$ acts as an effective topological deformation parameter that induces transitions between different winding-number classes.

\section{Topological Phase Structure for $F^{\alpha \beta } F^{\gamma \lambda } R_{\alpha \gamma } R_{\beta \lambda }$ Coupling }\label{sec5}

As the first step, we construct the off-shell free energy and the vector field component $\phi^{r_h}$. These are given by

\begin{equation}\label{fshell}
\begin{aligned}
\mathcal{F} = \mathcal{H} - S/\tau
&= \frac{4\pi P r_h^3}{3} + \frac{r_h}{2} + \frac{Q^2}{8r_h} \\
&\quad + \epsilon \Bigg(
  \frac{64\pi^2 P^2 Q^2}{r_h}
  + \frac{2\pi P Q^4}{5 r_h^5}
  - \frac{Q^6}{36 r_h^9}
  + \frac{4\pi P Q^2}{r_h^3}
  + \frac{Q^4}{8 r_h^7}
\Bigg) \\
&\quad - \frac{1}{\tau}
\Bigg(
  \pi r_h^2
  + \frac{16\epsilon \pi^2 P Q^2}{r_h^2}
  + \frac{\epsilon \pi Q^4}{2 r_h^6}
\Bigg),
\end{aligned}
\end{equation}

and

\begin{equation}\label{dfshell}
\begin{aligned}
\phi^{r_h}
&= \frac{\partial \mathcal{F}}{\partial r_h} \\
&= 4\pi P r_h^2 - \frac{Q^2}{8r_h^2} + \frac{1}{2} \\
&\quad + \epsilon
\Bigg(
  -\frac{64\pi^2 P^2 Q^2}{r_h^2}
  - \frac{12\pi P Q^2}{r_h^4}
  - \frac{2\pi P Q^4}{r_h^6}
  - \frac{7Q^4}{8 r_h^8}
  + \frac{Q^6}{4 r_h^{10}}
\Bigg) \\
&\quad - \frac{1}{\tau}
\Bigg(
  2\pi r_h
  - \frac{32\epsilon\pi^2 P Q^2}{r_h^3}
  - \frac{3\epsilon\pi Q^4}{r_h^7}
\Bigg).
\end{aligned}
\end{equation}

\begin{figure}[H]
    \centering
    \includegraphics[width=6cm]{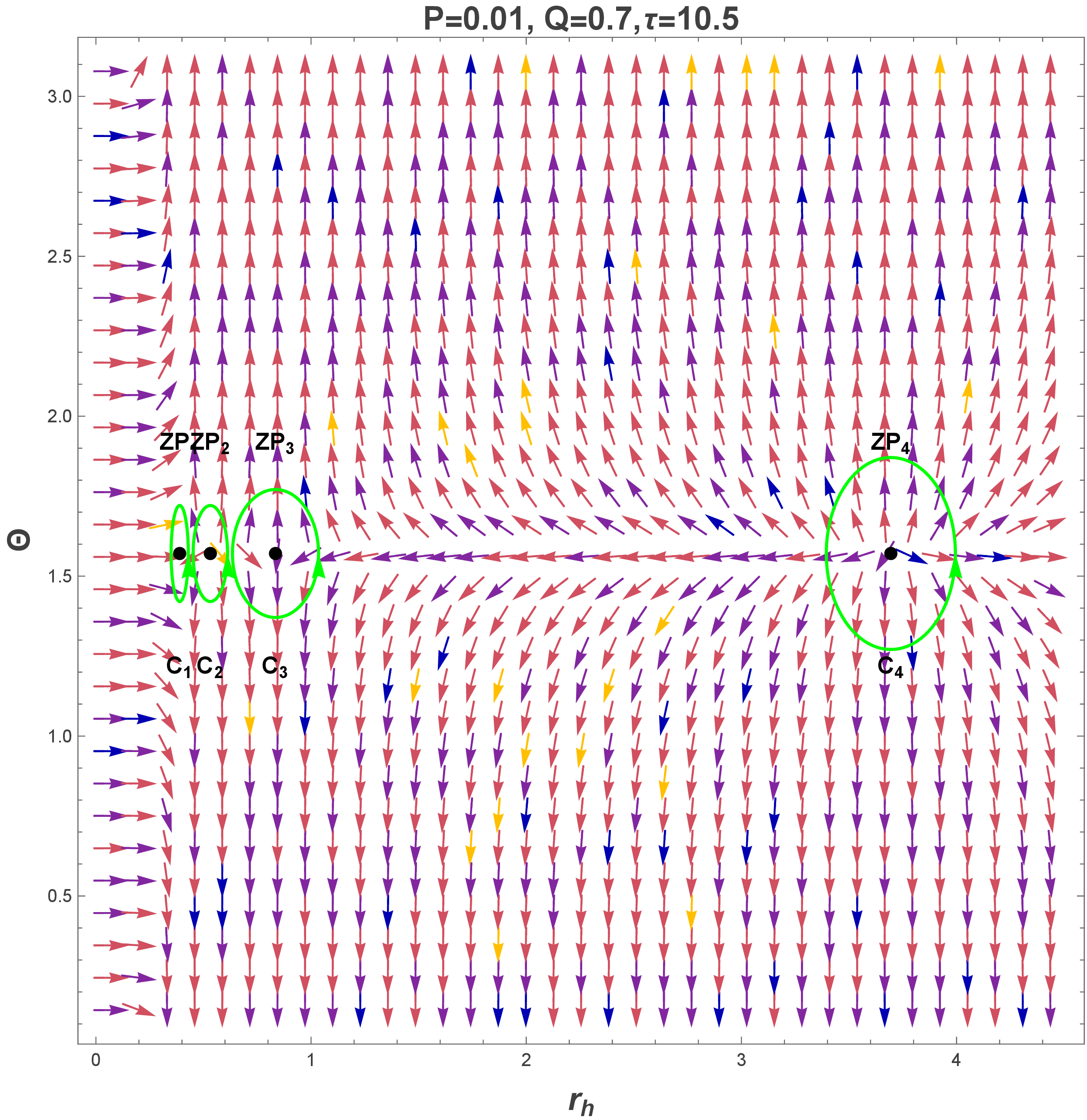}
    \includegraphics[width=6cm]{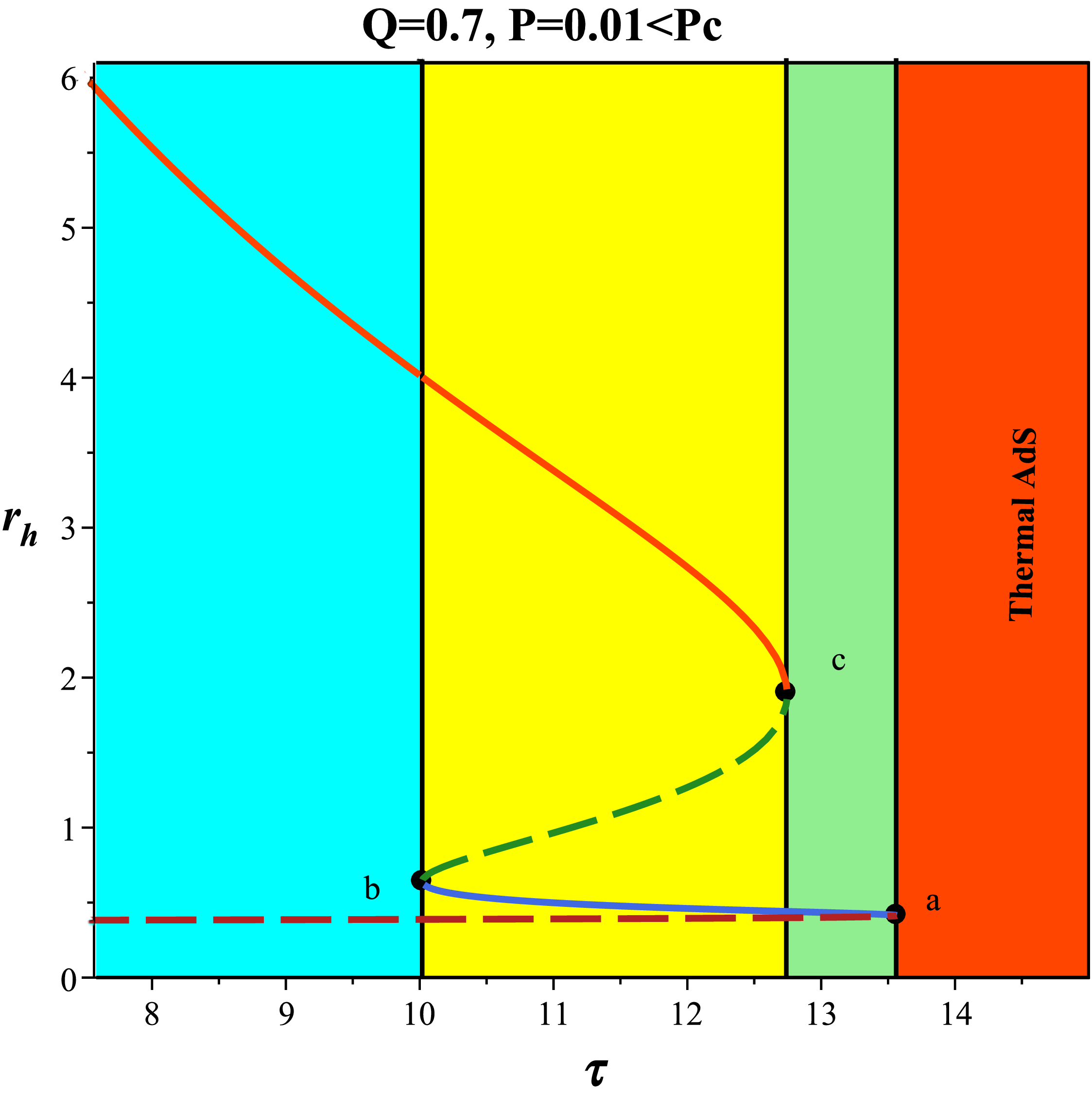}
    \caption{(a) Unit vector field for $P = 0.01$, $Q = 0.7$, and $\tau = 10.5$. The direction of the vector field at the boundaries places the system in the $W^{0-}$ topological class. (b) $r_h$--$\tau$ diagram for $Q = 0.7$ and $P = 0.01 < P_c$, illustrating the global thermodynamic behavior. The colored regions indicate different phases, with the total winding number $W = 0$ in each region.}
    \label{fig:v1rh1}
\end{figure}

The topological approach yields behavior consistent with conventional thermodynamic analysis. For instance, for $Q = 0.7$ and $P = 0.01 < P_c \approx 0.026$, the system exhibits four topological defects. A general Hawking--Page pattern always requires an even number of defects. In this case, two intermediate defects annihilate, leaving only the innermost unstable and outermost stable black holes. The behavior of the unit vector field at the boundaries, shown in the left panel of Fig.~\ref{fig:v1rh1}, places the system in the $W^{0-}$ topological class.

To locate these four defects, we first consider the critical temperature for $Q = 0.7$ and $\epsilon = 0.001$, which is $T_c \approx 0.123$. The corresponding inverse critical temperature is $\tau_c \approx 8.103$. By investigating $\tau$ as a function of $r_h$, one can determine the domain where the defects can appear. Van der Waals-type branches emerge as the temperature decreases (i.e., as $\tau$ increases), whereas the Hawking--Page branches appear as the temperature increases (as $\tau$ decreases). Since the system is hybrid, the van der Waals oscillations become negligible at higher temperatures, leaving the global Hawking--Page pattern dominant. Thus, to observe van der Waals behavior, one must decrease the temperature (increase $\tau$), and four defects are expected for $\tau > \tau_c$. The left panel of Fig.~\ref{fig:v1rh1} is plotted for $\tau = 10.5$, which lies in this regime.  

The right panel displays the global behavior of the system in the $r_h$--$\tau$ plane. Solving $\phi^{r_h}=0$ for $\tau$ gives
\begin{equation}
\tau = \frac{2\pi r_h - \frac{32\epsilon\pi^2 P Q^2}{r_h^3} - \frac{3\epsilon\pi Q^4}{r_h^7}}{\epsilon\left(-\frac{64\pi^2 P^2 Q^2}{r_h^2} - \frac{12\pi P Q^2}{r_h^4} - \frac{2\pi P Q^4}{r_h^6} - \frac{7Q^4}{8r_h^8} + \frac{Q^6}{4r_h^{10}}\right) + 4\pi P r_h^2 - \frac{Q^2}{8r_h^2} + \frac{1}{2}}.
\end{equation}

In each colored region, the total winding number $W$ vanishes. As $\tau$ increases, the annihilation point $a$ appears, followed by the generation point $b$, and then the annihilation point $c$. For $\tau > \tau_a = 13.56$, only thermal AdS space exists. Between $\tau_a$ and $\tau_c$, two black hole states coexist: an ultra-small unstable black hole and a small stable black hole. In the interval $\tau_b < \tau < \tau_c$, four branches coexist. For $\tau < \tau_b$, two branches are present: an ultra-small unstable black hole and a large stable branch. In the limits $r_h \to 0$ and $r_h \to \infty$, the inverse temperature $\beta$ tends to zero.

\begin{figure}[H]
    \centering
    \includegraphics[width=0.38\textwidth]{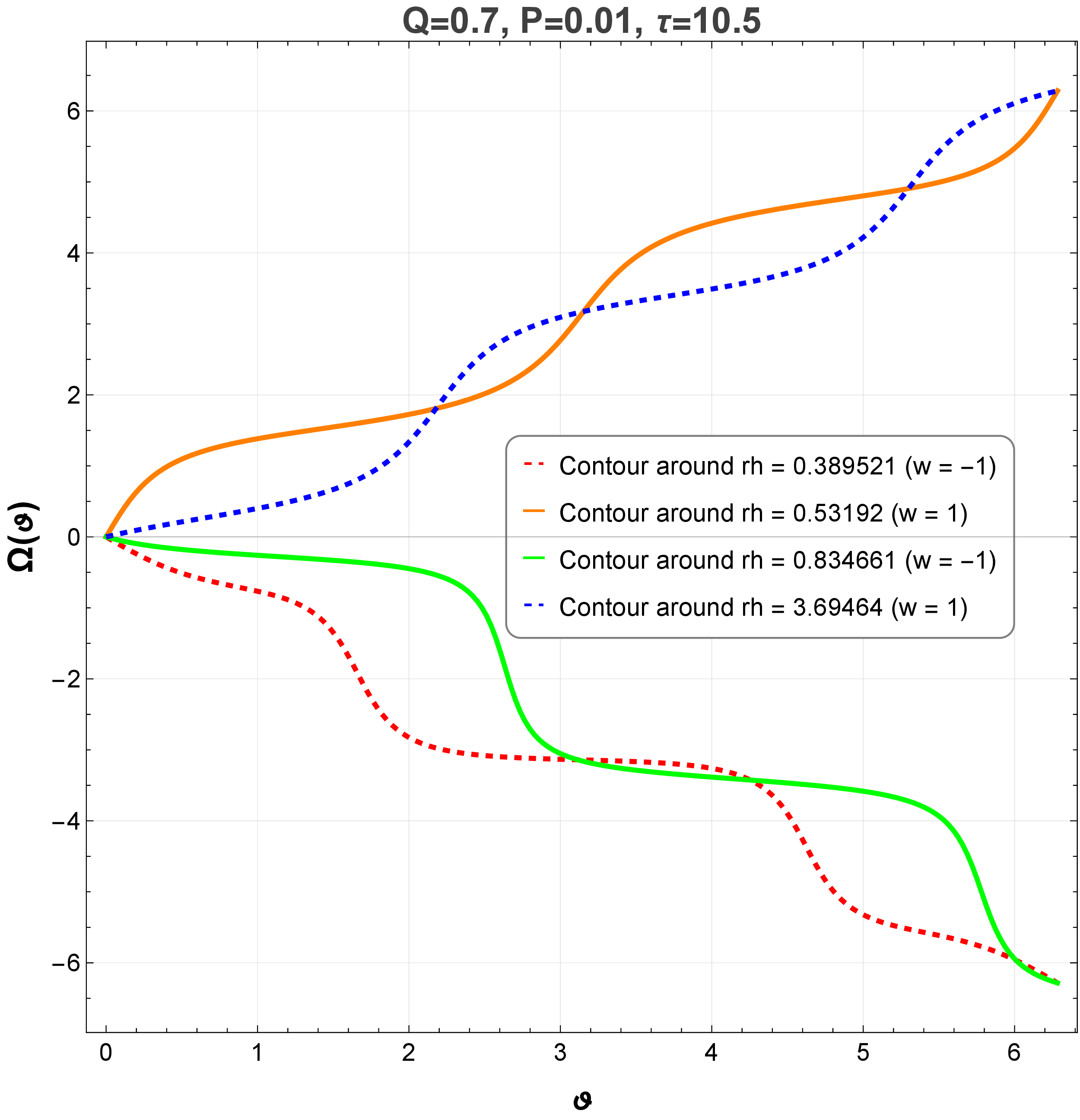}
    \quad
    \includegraphics[width=0.38\textwidth,height=0.31\textheight]{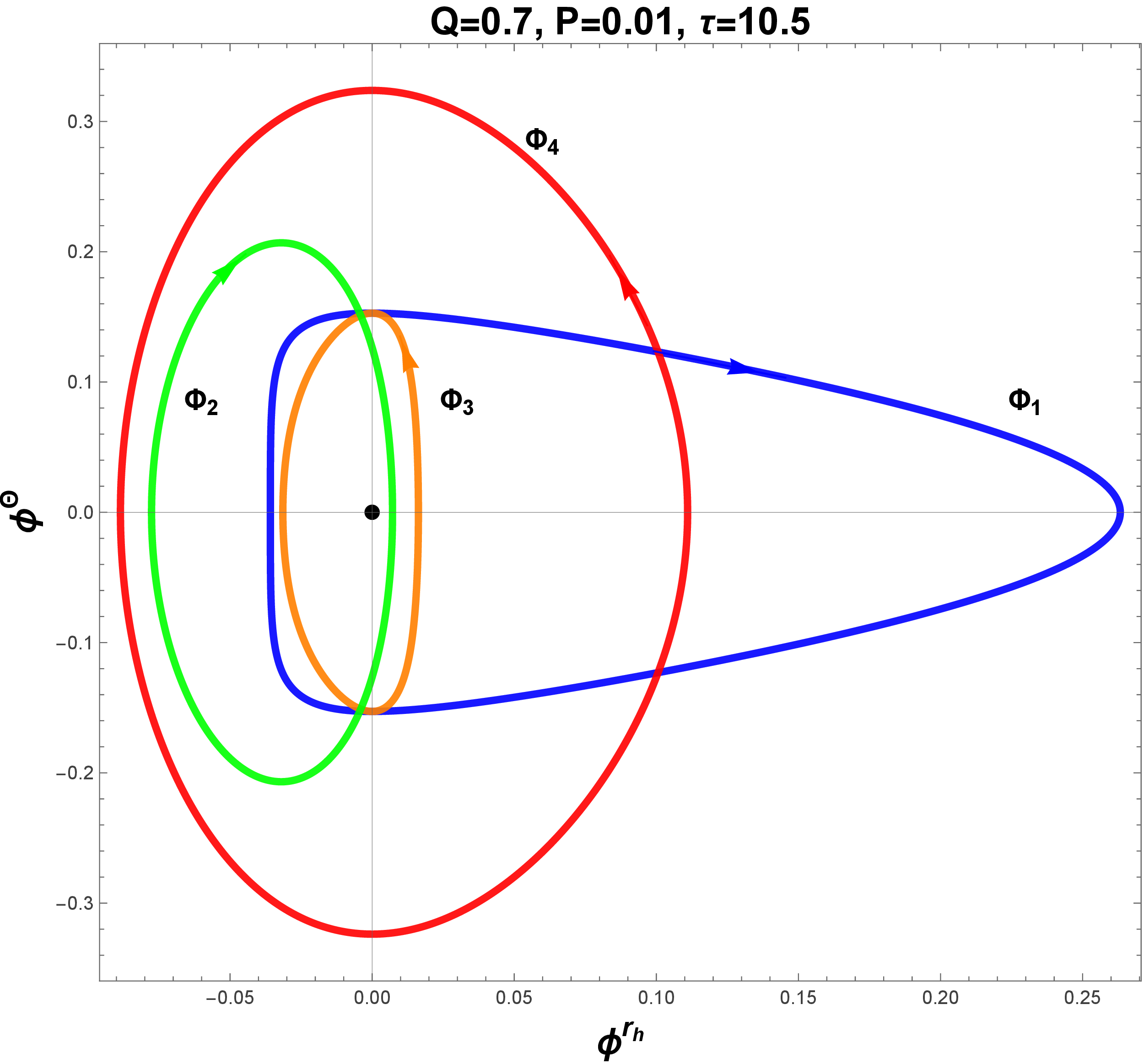}
    \caption{(a) Deflection angle $\Omega(\vartheta)$ for a closed loop enclosing four defects at $\mathrm{ZP}_1 = 0.389$, $\mathrm{ZP}_2 = 0.531$, $\mathrm{ZP}_3 = 0.834$, and $\mathrm{ZP}_4 = 3.694$, for $Q = 0.7$, $P = 0.01$, and $\tau = 10.5$. The total deflection after one full circuit yields the winding numbers $w = -1, +1, -1, +1$, respectively. (b) Contour mapping in the $(\phi^{r_h}, \phi^\Theta)$ plane for the same parameter values. Each closed curve corresponds to a contour encircling a defect in the $(r_h, \Theta)$-plane. Counterclockwise traversal corresponds to $w = +1$ (stable branch), while clockwise traversal corresponds to $w = -1$ (unstable branch). The four defects in the $r_h$--$\Theta$ plane are mapped to the origin in the $\phi$-plane.}
    \label{fig:omega1con1}
\end{figure}

To verify the winding numbers associated with each defect in Fig.~\ref{fig:v1rh1}, we use the deflection angle relation from Eq.~\eqref{winding} to plot the deflection angle diagram for $0 < \vartheta < 2\pi$. The left panel of Fig.~\ref{fig:omega1con1} illustrates this result. The figure shows the deflection angle plot for contours enclosing each defect located at $\mathrm{ZP}_1 = 0.389$, $\mathrm{ZP}_2 = 0.531$, $\mathrm{ZP}_3 = 0.834$, and $\mathrm{ZP}_4 = 3.694$, for $Q = 0.7$, $P = 0.01$, and $\tau = 10.5$. The winding numbers are found to be $w = -1, +1, -1, +1$, respectively.

A further consistency check can be performed by mapping contours around the zero points in the $r_h$--$\Theta$ plane to the $(\phi^{r_h}, \phi^\Theta)$ vector-field plane. Consider the closed contours $C_i$ in the $(r_h, \Theta)$-plane that encircle each zero point of the vector field $\phi$. Under the mapping $\phi: (r_h, \Theta) \mapsto (\phi^{r_h}, \phi^\Theta)$, each contour $C_i$ maps to a corresponding closed curve $\Phi_i$ in the $(\phi^{r_h}, \phi^\Theta)$-plane. The winding number is then determined by the number of times $\Phi_i$ winds around the origin. By convention, the contour $C_i$ in the $r_h$--$\Theta$ plane is traversed counterclockwise. If the image contour $\Phi_i$ in the $(\phi^{r_h}, \phi^\Theta)$ plane is also traversed counterclockwise, the winding number is positive ($w_i = +1$), indicating a stable black hole state. Conversely, if $\Phi_i$ is traversed clockwise, the winding number is negative ($w_i = -1$), corresponding to an unstable black hole state. The right panel of Fig.~\ref{fig:omega1con1} illustrates this mapping for the same parameter values.

\begin{figure}[H]
    \centering
    \includegraphics[width=0.38\textwidth]{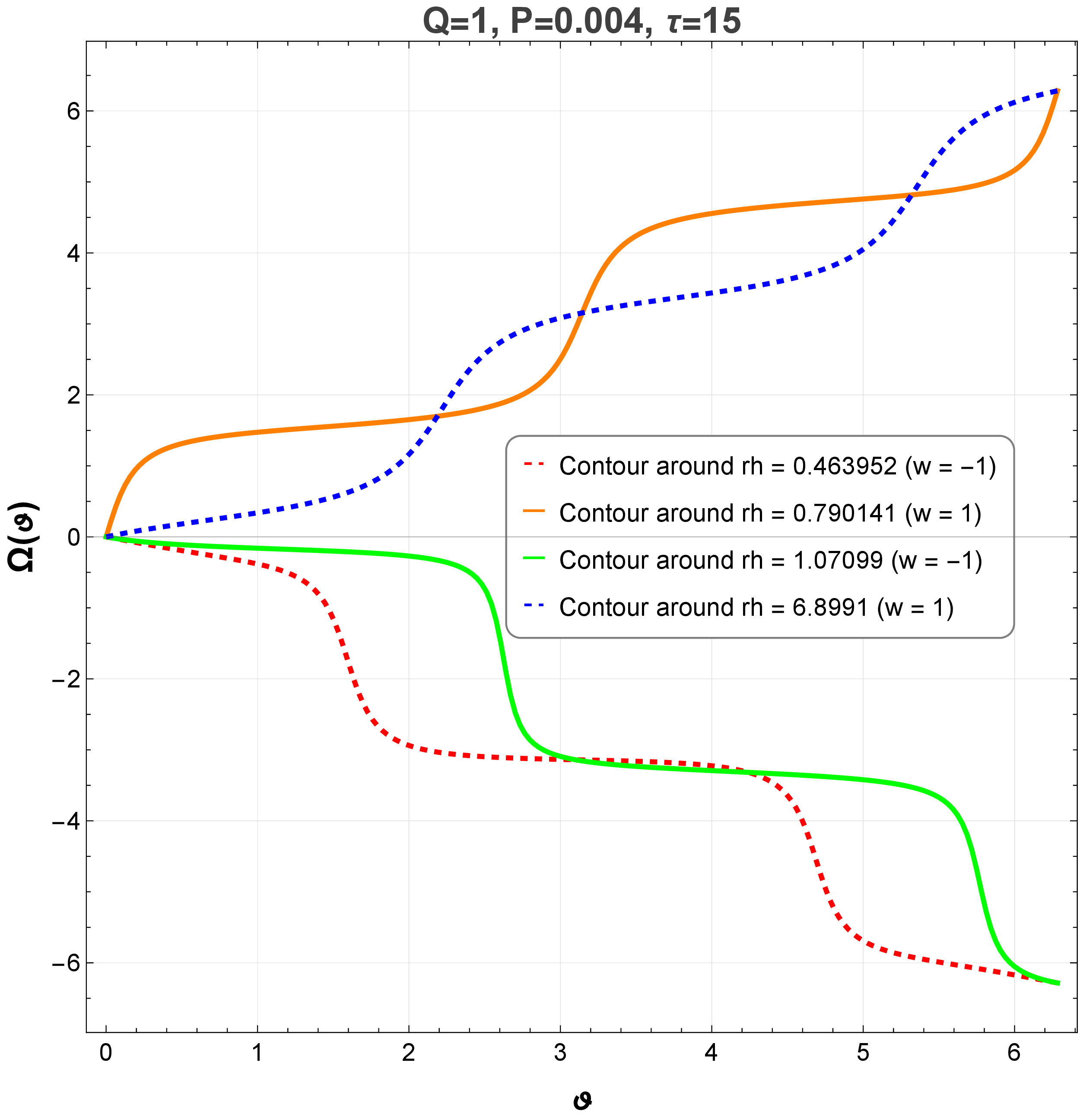}
    \quad
    \includegraphics[width=0.38\textwidth,height=0.31\textheight]{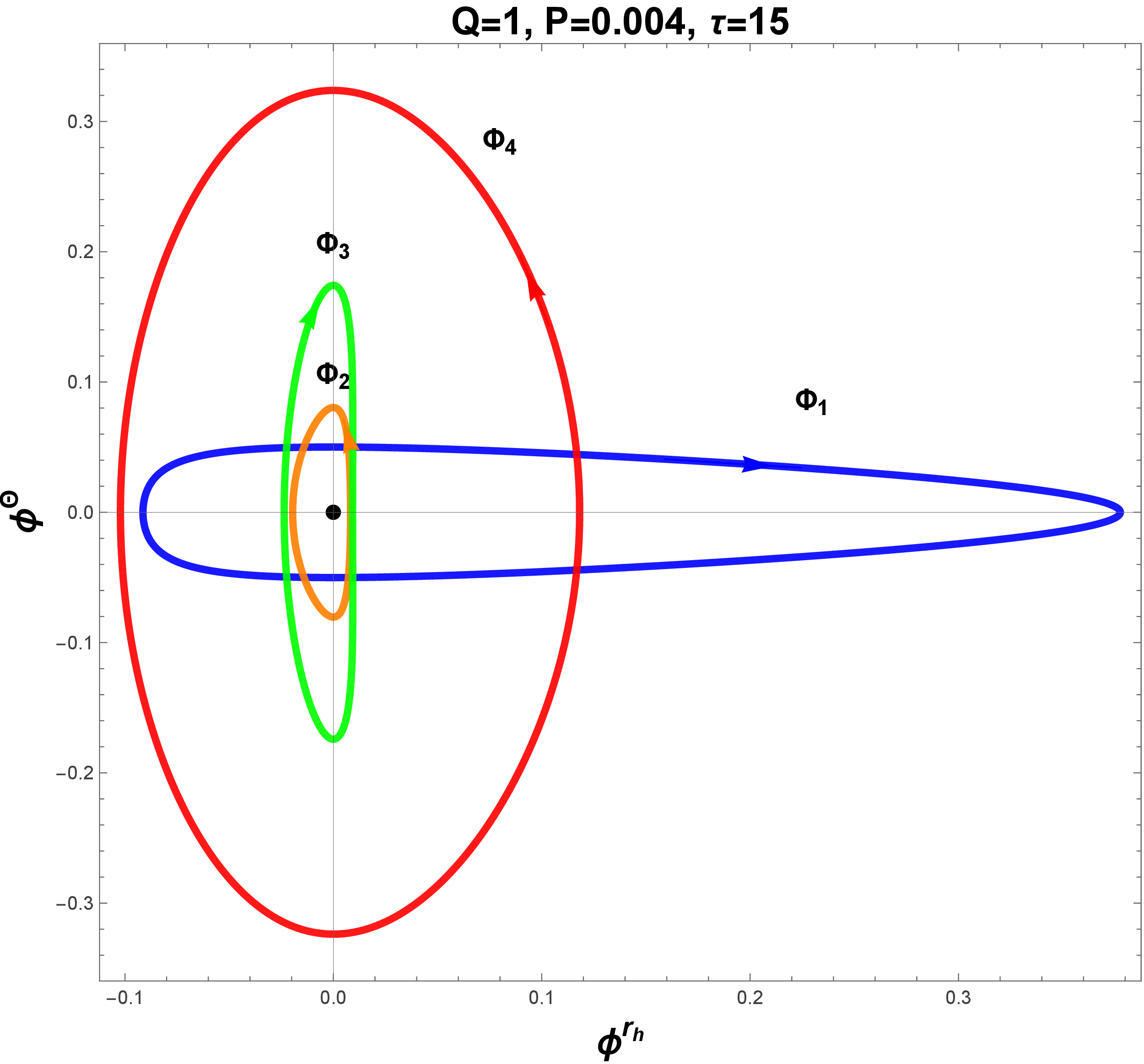}
    \caption{(a) Deflection angle $\Omega(\vartheta)$ for a closed loop enclosing four defects located at $\mathrm{ZP}_1 = 0.4639$, $\mathrm{ZP}_2 = 0.7901$, $\mathrm{ZP}_3 = 1.070$, and $\mathrm{ZP}_4 = 6.899$, for $Q = 1$, $P = 0.004$, and $\tau = 15$. The total deflection after one full circuit yields the winding numbers $w = -1, +1, -1, +1$, respectively. (b) Contour mapping in the $(\phi^{r_h}, \phi^\Theta)$ plane for the same parameter values. Each closed curve corresponds to a contour encircling a defect in the $(r_h, \Theta)$-plane. Counterclockwise traversal corresponds to $w = +1$ (stable branch), while clockwise traversal corresponds to $w = -1$ (unstable branch). The four defects in the $r_h$--$\Theta$ plane are mapped to the origin in the $\phi$-plane.}
    \label{fig:omega2con2}
\end{figure}

The deflection angle method and contour mapping method have been used for parameter values $Q=1$, $P=0.004$, $\tau=15$, and the result is illustrated in Fig.~\ref{fig:omega2con2}.

\begin{figure}[H]
    \centering
    \includegraphics[width=0.38\textwidth]{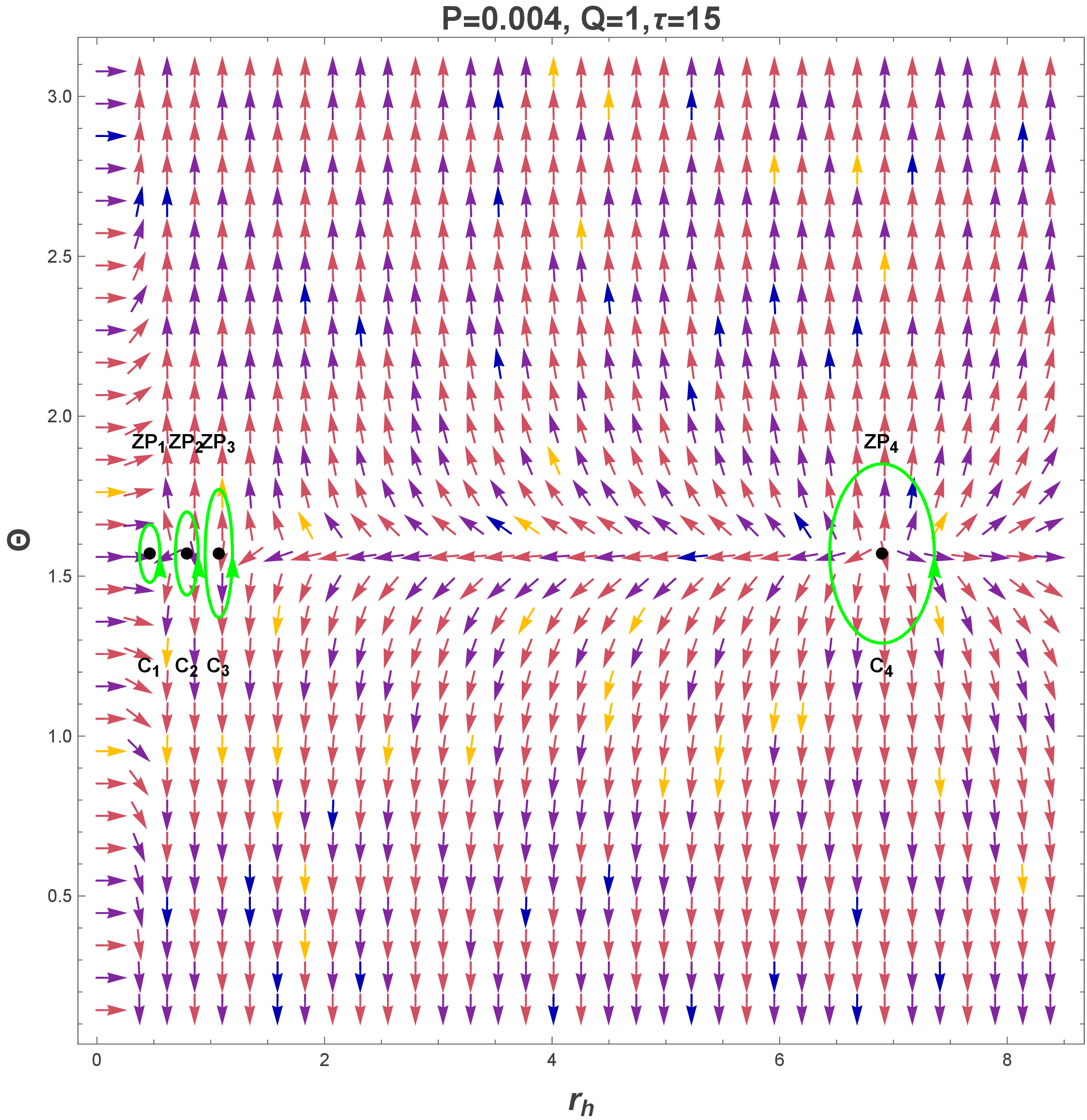}
    \quad
    \includegraphics[width=0.38\textwidth]{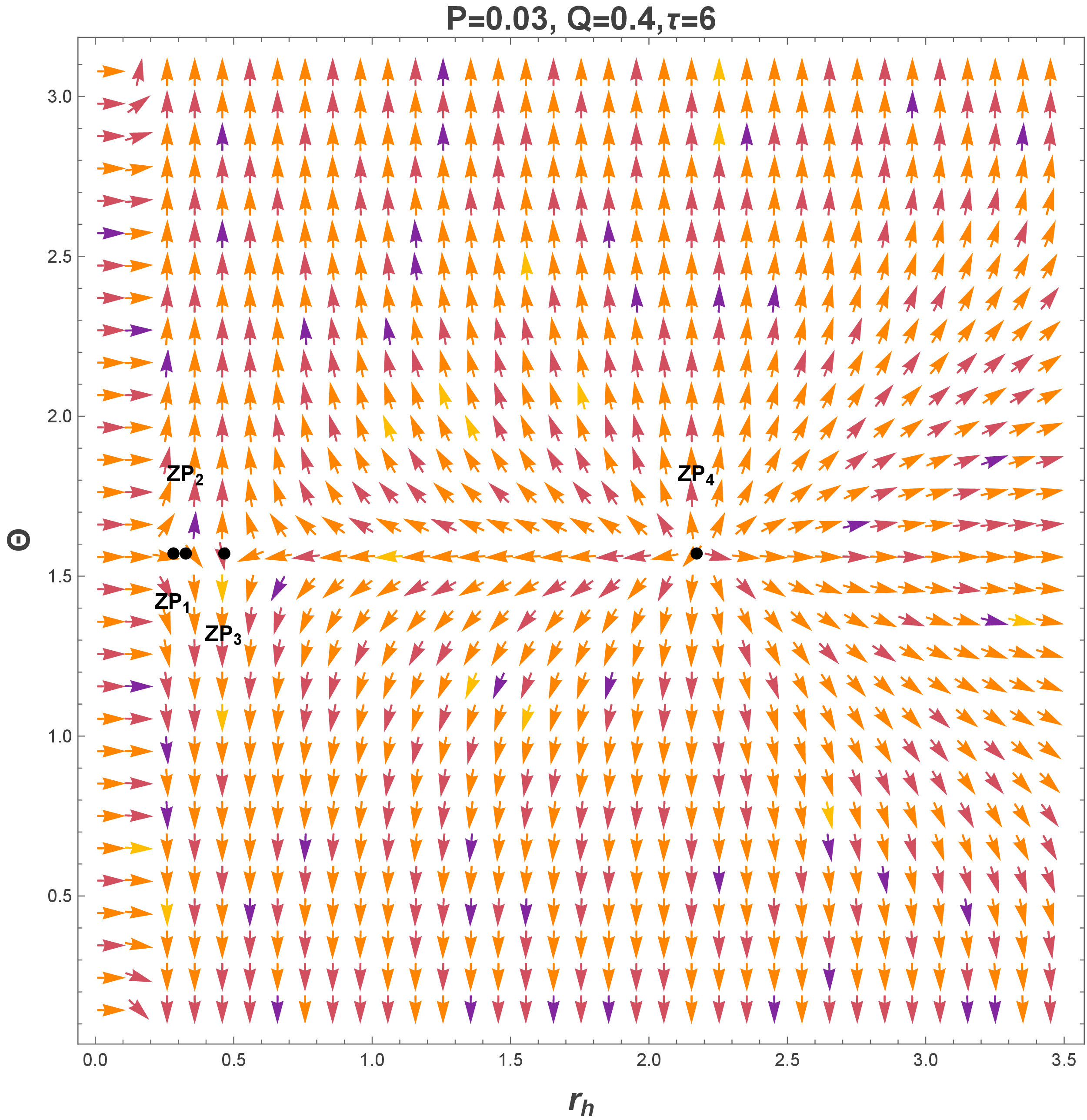}\\
    \includegraphics[width=0.38\textwidth]{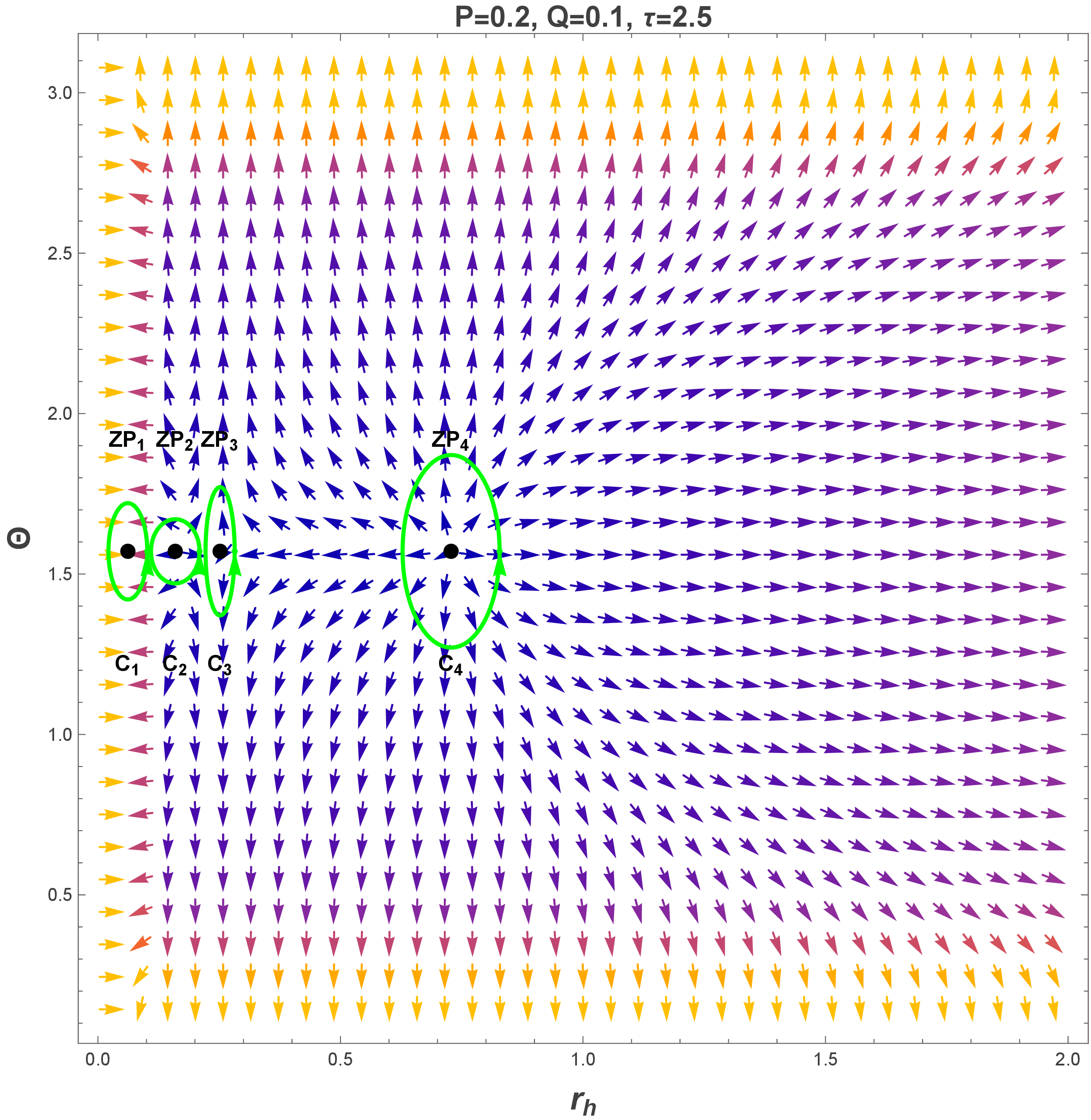}
    \quad
    \includegraphics[width=0.38\textwidth]{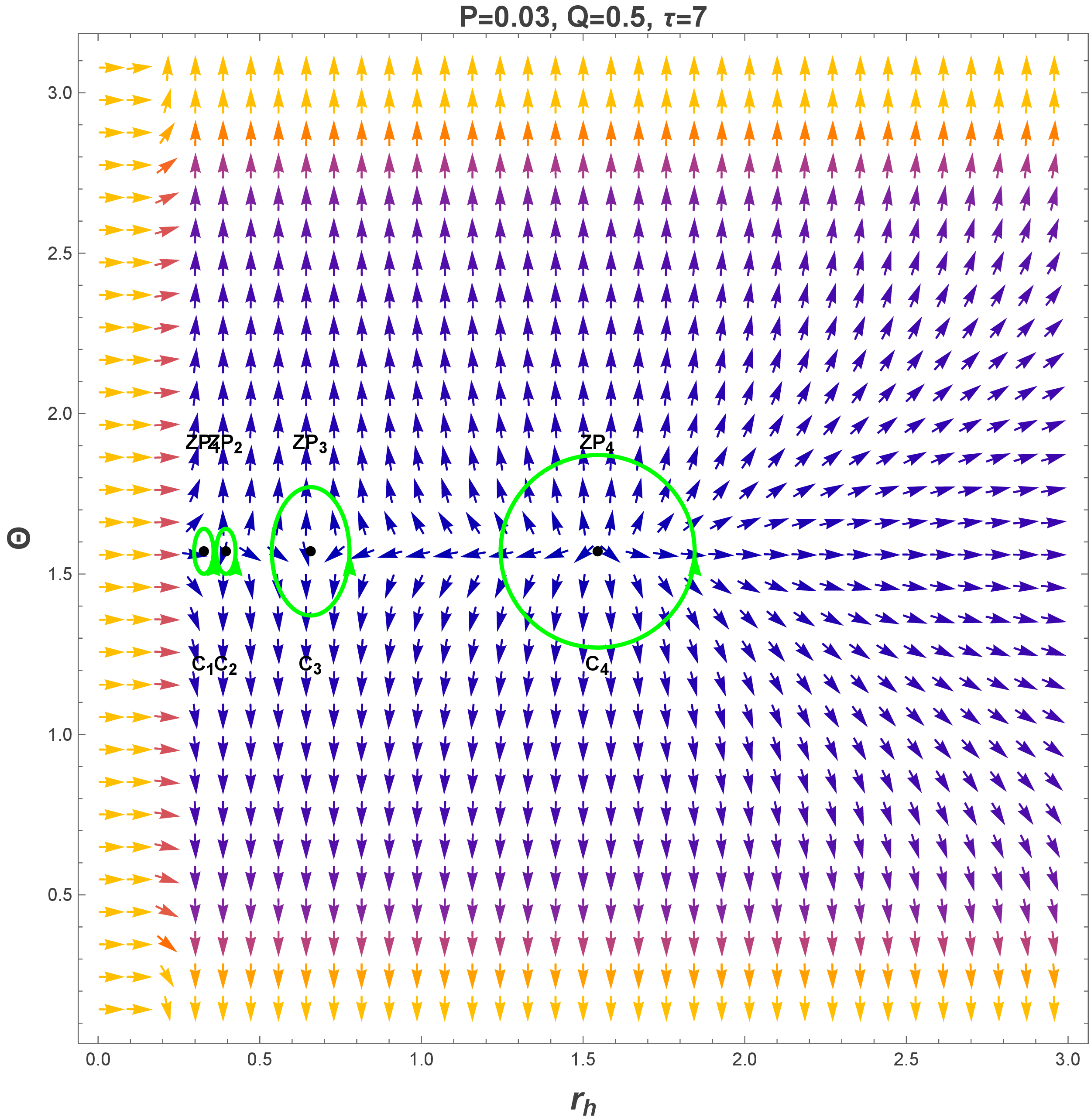}
    \caption{(a) Unit vector field for $Q = 1$, $P = 0.004 < P_c = 0.013$, and $\tau = 15$, showing four topological defects. (b) Unit vector field for $Q = 0.4$, $P = 0.04 < P_c = 0.079$, and $\tau = 6$; the defects are closely spaced, and the conventional loops encircling each zero point are omitted for clarity. (c) Unit vector field for $Q = 0.1$, $P = 0.2 < P_c = 0.509$, and $\tau = 2.5$, exhibiting four distinct defects. (d) Unit vector field for $Q = 0.5$, $P = 0.03 < P_c = 0.052$, and $\tau = 7$, again displaying four topological defects. These plots further confirm that the system belongs to the $W^{0-}$ class.}
    \label{fig:v2v3v4v5}
\end{figure}

For further confirmation, we illustrate the unit vector field for different parameter values in Fig.~\ref{fig:v2v3v4v5}. Panels (a), (b), (c), and (d) correspond to the following parameter sets, respectively: $Q = 1$, $P = 0.004 < P_c = 0.013$, $\tau = 15$; $Q = 0.4$, $P = 0.04 < P_c = 0.079$, $\tau = 6$; $Q = 0.1$, $P = 0.2 < P_c = 0.509$, $\tau = 2.5$; and $Q = 0.5$, $P = 0.03 < P_c = 0.052$, $\tau = 7$.

In panel (b), the defects are very closely spaced; therefore, we omit the conventional loops encircling each zero point for clarity.

The boundary behavior of the unit vector fields consistently places the system in the $W^{0-}$ topological classification across all cases.
The topological formalism reveals that the nonminimal coupling $\epsilon F^{\alpha\beta}F^{\gamma\lambda}R_{\alpha\gamma}R_{\beta\lambda}$ acts as a topological deformation parameter, capable of inducing transitions between distinct winding number classes. Within the extended topological classification, standard classes include $W^{1-}$ (Schwarzschild), $W^{0+}$ (Reissner--Nordstr\"om), $W^{0-}$ (Schwarzschild-AdS), and $W^{1+}$ (Reissner--Nordstr\"om-AdS). Despite exhibiting van der Waals oscillations, the system remains in $W^{0-}$ because the asymptotic behavior is controlled by the Hawking--Page skeleton, which is preserved by $\epsilon$ even when $Q \neq 0$.  

One may ask how a first-order perturbative solution in $\epsilon$ can lead to a genuine topological phase transition. Since topology is discrete, should the topological class not remain unchanged for sufficiently small $\epsilon$?

The topological classification is based entirely on the asymptotic behavior of the inverse temperature $\beta(r_h) = 1/T(r_h)$ in the limits $r_h \to r_m$ (the minimal horizon radius) and $r_h \to \infty$. The topological number $W$ is a discrete invariant that depends only on these boundary limits, and not on the magnitude of any coupling parameter. The key mechanism lies in the behavior of the temperature as $r_h \to 0$. From Eq.~(35):

\[
T(r_h) = \underbrace{2Pr_h + \frac{1}{4\pi r_h} - \frac{Q^2}{16\pi r_h^3}}_{\epsilon=0 \text{ terms}}
+ \epsilon\underbrace{\left(-\frac{2PQ^2}{r_h^5} + \frac{PQ^4}{r_h^7} - \frac{Q^4}{16\pi r_h^9} + \frac{Q^6}{32\pi r_h^{11}}\right)}_{\text{dominant as } r_h \to 0}.
\]

For $\epsilon = 0$, as $r_h \to 0$, the term $-Q^2/(16\pi r_h^3)$ drives $T \to -\infty$, so $T(r_m) = 0$ determines a finite minimal radius with $\beta(r_m) = \infty$. For any $\epsilon \neq 0$, the term $+\epsilon Q^6/(32\pi r_h^{11})$ dominates as $r_h \to 0$, driving $T \to +\infty$. Therefore, $\beta(r_m) = 0$ for \emph{any} non-zero $\epsilon$.

The perturbative expansion remains well-controlled in the region of interest. As shown in Appendix C, the dimensionless expansion parameter is $\xi = \epsilon/\ell^4 \sim 5.09 \times 10^{-5}$ for our primary value $\epsilon = 0.001$. The critical point shifts are at the $0.26\%$ level (Table~5), and the linearity tests (Table~6) confirm the validity of the first-order approximation.


\section{Conclusions}
\label{sec6}

In this work, we have investigated the conventional and topological phase transitions of a four-dimensional asymptotically AdS black hole with a non-minimal coupling term of the form $F^{\alpha\beta}F^{\gamma\lambda}R_{\alpha\gamma}R_{\beta\lambda}$. Due to the higher-derivative nature of this interaction, perturbative black hole solutions were constructed to first order in the coupling parameter $\epsilon$, explicitly deriving the metric function, gauge potential, and all associated thermodynamic quantities. We verified the first law of thermodynamics and the Smarr relation to first order in $\epsilon$, confirming the internal consistency of the perturbation scheme.

The conventional thermodynamic analysis revealed that this system exhibits a remarkable hybrid behavior: a global Hawking--Page pattern coexists with a van der Waals-type first-order phase transition embedded within the lower branch of the free energy. The heat capacity displays alternating stable and unstable branches, confirming the simultaneous presence of both transition mechanisms. Unlike standard RN-AdS black holes, which exhibit only van der Waals behavior in the canonical ensemble, or Schwarzschild-AdS black holes, which exhibit only the Hawking-Page transition, the non-minimally coupled system studied here simultaneously displays both phenomena. This distinguishes it from previously studied systems in the extended topological classification.

The most striking result emerged from the topological analysis based on Duan's $\phi$-mapping theory. While the $\epsilon = 0$ limit corresponds to the standard Reissner--Nordstr\"om--AdS black hole belonging to the $W^{1+}$ topological class with $W = +1$, switching on the non-minimal coupling fundamentally transforms the topology to the $W^{0-}$ class with $W = 0$. This transition occurs because the topological class is determined solely by the asymptotic behavior of $\beta(r_h)$ as $r_h \to 0$ and $r_h \to \infty$. The $\epsilon$-terms in the temperature introduce a $+\epsilon Q^6/(32\pi r_h^{11})$ divergence as $r_h \to 0$, flipping $\beta(r_m)$ from $\infty$ to $0$ for any $\epsilon \neq 0$. The perturbative expansion remains well-controlled ($\xi = \epsilon/\ell^4 \sim 5.09 \times 10^{-5}$ for $\epsilon = 0.001$), ensuring the robustness of the topological classification. The coupling $\epsilon$ thus acts as a topological deformation parameter capable of altering the universal classification of the black hole thermodynamic system, even at the perturbative level.

The hybrid nature of this system may also have implications for the AdS/CFT correspondence. The Hawking--Page transition is generally associated with the confinement--deconfinement transition, whereas the van der Waals transition is analogous to a liquid--gas phase transition. Their coexistence therefore suggests the possibility of two distinct phase transitions in the dual CFT occurring at different temperatures: a small-to-intermediate black hole transition at lower temperatures, followed by a transition to the large black hole phase at higher temperatures.

In this work, we have focused on the canonical ensemble, where the electric charge $Q$ is held fixed. A natural and important extension is to analyze the system in the grand canonical ensemble, where the electric potential $\psi$ is fixed instead of $Q$. This analysis is left for future work.
\\

\vspace{1cm}
\noindent \textbf{Data Availability Statement:} No data were generated or analyzed in this study; therefore, data sharing is not applicable.

\appendix

\section{Explicit form of \(T^{(I)}_{\mu\nu}\)}
\label{app:A}

The energy-momentum tensor arising from the nonminimal coupling is
\begin{align}
	T^{(I)}_{\mu \nu }
	&= \tfrac{1}{2} F^{ \alpha \beta } F^{\gamma \lambda } g_{\mu \nu } R_{\alpha \gamma } R_{\beta \lambda }
	- 2 F^{\beta \gamma } F_{\nu }{}^{\alpha } R_{\alpha \gamma } R_{\mu \beta }
	- 2 F^{\beta \gamma } F_{\mu }{}^{\alpha } R_{\alpha \gamma } R_{\nu \beta }
	- F^{\alpha \beta } g_{\mu \nu } R_{\alpha \gamma } \nabla_{\beta }\nabla_{\lambda }F^{\gamma \lambda } \nonumber \\
	&\quad - F^{\alpha \beta } R_{\alpha \gamma } \nabla_{\beta }\nabla_{\mu }F_{\nu }{}^{\gamma }
	- F^{\alpha \beta } R_{\alpha \gamma } \nabla_{\beta }\nabla_{\nu }F_{\mu }{}^{\gamma }
	+ 2 F^{\alpha \beta } g_{\mu \nu } \nabla_{\beta }R_{\alpha }{}^{\lambda } \nabla_{\gamma }F^{\gamma }{}_{\lambda }
	- F_{\nu }{}^{\alpha } R_{\alpha \beta } \nabla_{\gamma }\nabla^{\gamma }F_{\mu }{}^{\beta } \nonumber \\
	&\quad - F_{\mu }{}^{\alpha } R_{\alpha \beta } \nabla_{\gamma }\nabla^{\gamma }F_{\nu }{}^{\beta }
	- F_{\mu }{}^{\alpha } F_{\nu }{}^{\beta } \nabla_{\gamma }\nabla^{\gamma }R_{\alpha \beta }
	- F_{\nu }{}^{\alpha } R_{\alpha \beta } \nabla_{\gamma }\nabla_{\mu }F^{\beta \gamma }
	- F^{\beta \gamma } F_{\nu }{}^{\alpha } \nabla_{\gamma }\nabla_{\mu }R_{\alpha \beta } \nonumber \\
	&\quad - F_{\mu }{}^{\alpha } R_{\alpha \beta } \nabla_{\gamma }\nabla_{\nu }F^{\beta \gamma }
	- F^{\beta \gamma } F_{\mu }{}^{\alpha } \nabla_{\gamma }\nabla_{\nu }R_{\alpha \beta }
	- 2 R_{\alpha \beta } \nabla_{\gamma }F_{\mu }{}^{\alpha } \nabla^{\gamma }F_{\nu }{}^{\beta }
	- 2 F_{\nu }{}^{\alpha } \nabla_{\gamma }F_{\mu }{}^{\beta } \nabla^{\gamma }R_{\alpha \beta } \nonumber \\
	&\quad - 2 F_{\mu }{}^{\alpha } \nabla_{\gamma }F_{\nu }{}^{\beta } \nabla^{\gamma }R_{\alpha \beta }
	- g_{\mu \nu } R_{\alpha \beta } \nabla_{\gamma }F^{\beta \lambda } \nabla_{\lambda }F^{\alpha \gamma }
	- g_{\mu \nu } R_{\alpha \beta } \nabla_{\gamma }F^{\alpha \gamma } \nabla_{\lambda }F^{\beta \lambda }
	+ 2 F^{\alpha \beta } g_{\mu \nu } \nabla_{\alpha }F^{\gamma \lambda } \nabla_{\lambda }R_{\beta \gamma } \nonumber \\
	&\quad - F^{\alpha \beta } g_{\mu \nu } R_{\alpha \gamma } \nabla_{\lambda }\nabla_{\beta }F^{\gamma \lambda }
	- F^{\alpha \beta } F^{\gamma \lambda } g_{\mu \nu } \nabla_{\lambda }\nabla_{\beta }R_{\alpha \gamma }
	- F_{\nu }{}^{\alpha } \nabla_{\gamma }R_{\alpha }{}^{\beta } \nabla_{\mu }F_{\beta }{}^{\gamma }
	- R_{\alpha \beta } \nabla_{\gamma }F_{\nu }{}^{\alpha } \nabla_{\mu }F^{\beta \gamma } \nonumber \\
	&\quad - R_{\alpha \beta } \nabla_{\gamma }F^{\beta \gamma } \nabla_{\mu }F_{\nu }{}^{\alpha }
	- F^{\alpha \beta } \nabla_{\beta }R_{\alpha }{}^{\gamma } \nabla_{\mu }F_{\nu \gamma }
	+ F_{\nu }{}^{\alpha } \nabla_{\beta }F^{\beta \gamma } \nabla_{\mu }R_{\alpha \gamma }
	+ F^{\alpha \beta } \nabla_{\alpha }F_{\nu }{}^{\gamma } \nabla_{\mu }R_{\beta \gamma } \nonumber \\
	&\quad - F_{\mu }{}^{\alpha } \nabla_{\gamma }R_{\alpha }{}^{\beta } \nabla_{\nu }F_{\beta }{}^{\gamma }
	- R_{\alpha \beta } \nabla_{\gamma }F_{\mu }{}^{\alpha } \nabla_{\nu }F^{\beta \gamma }
	- R_{\alpha \beta } \nabla_{\gamma }F^{\beta \gamma } \nabla_{\nu }F_{\mu }{}^{\alpha }
	- F^{\alpha \beta } \nabla_{\beta }R_{\alpha }{}^{\gamma } \nabla_{\nu }F_{\mu \gamma } \nonumber \\
	&\quad + F_{\mu }{}^{\alpha } \nabla_{\beta }F^{\beta \gamma } \nabla_{\nu }R_{\alpha \gamma }
	+ F^{\alpha \beta } \nabla_{\alpha }F_{\mu }{}^{\gamma } \nabla_{\nu }R_{\beta \gamma }.
\end{align}

This tensor contributes to the gravitational field equations and is obtained from the variation of the nonminimal interaction term $ \epsilon F^{\alpha\beta}F^{\gamma\lambda}R_{\alpha\gamma}R_{\beta\lambda}$ with respect to the metric \(g^{\mu\nu}\).

\section{Explicit Form of the Field Equations at First Order}
\label{app:B}

In this appendix, we present the explicit form of the \(tt\) and \(rr\) components of the gravitational field equations and Maxwell equations at first order in the nonminimal coupling parameter \(\epsilon\). These equations were used in Sec.~\ref{sec2} to derive the perturbative solutions for the metric functions and gauge potential.

The \(tt\) component of Eq.~(\ref{EOM1}) is:
\begin{equation}\label{tteq}
4 r f_0' + 4 f_0 + \kappa \alpha r^2 e^{2H_0} h_0'^2 + 4\Lambda r^2  - \epsilon B_1(r) = 0,
\end{equation}
where \(B_1(r)\) is defined as:
\begin{equation}
\begin{aligned}
B_1(r) &= 3 \mathrm{e}^{2H(r)} \kappa r^2 h'(r)^2 f''(r)^2 \\
&\quad + \mathrm{e}^{2H(r)} \kappa f'(r)^2 h'(r) \Big[ 4 r \big(-2 + 3 r H'(r)\big) h''(r) + h'(r) \big( 8 - 28 r H'(r) + 35 r^2 H'(r)^2 + 10 r^2 H''(r) \big) \Big] \\
&\quad - 2 r f'(r) \Big[ 2 \mathrm{e}^{2H(r)} \kappa r h'(r) f''(r) h''(r) + \mathrm{e}^{2H(r)} \kappa h'(r)^2 \big( (-2 + 8 r H'(r)) f''(r) + r f^{(3)}(r) \big) \Big] \\
&\quad - 4 f(r) \Bigg[ 2 \mathrm{e}^{2H(r)} \kappa r^2 f''(r) h''(r)^2 \\
&\qquad - 2 \mathrm{e}^{2H(r)} \kappa r h'(r) \Big( -2 r h''(r) f^{(3)}(r) + f''(r) \big( (-7 + 2 r H'(r)) h''(r) - r h^{(3)}(r) \big) \Big) \\
&\qquad + \mathrm{e}^{2H(r)} \kappa f'(r) \Big( 2 r (2 - 3 r H'(r)) h''(r)^2 \\
&\qquad\quad - 2 h'(r) \Big( h''(r) \big( -2 + 6 r H'(r) + 7 r^2 H'(r)^2 + 11 r^2 H''(r) \big) + r (-2 + 3 r H'(r)) h^{(3)}(r) \Big) \\
&\qquad\quad + r h'(r)^2 \big( -27 H'(r)^2 + 7 r H'(r)^3 - 12 H''(r) - 27 r H'(r) H''(r) - 8 r H^{(3)}(r) \big) \Big) \\
&\qquad - \mathrm{e}^{2H(r)} \kappa h'(r)^2 \Big( 3 f''(r) \big( -1 - 2 r H'(r) + 3 r^2 H'(r)^2 + r^2 H''(r) \big) \\
&\qquad\quad - r \big( (5 + r H'(r)) f^{(3)}(r) + r f^{(4)}(r) \big) \Big) \Bigg] \\
&\quad - 4 \mathrm{e}^{2H(r)} \kappa f(r)^2 \Bigg[ 4 r h''(r)^2 \big( -H'(r) + r H'(r)^2 - r H''(r) \big) \\
&\qquad + 4 h'(r) \Big( 4 r^2 H'(r)^3 h''(r) + r H'(r)^2 \big( -h''(r) + r h^{(3)}(r) \big) \\
&\qquad\quad - H'(r) \big( 2 h''(r) + r h^{(3)}(r) \big) - r \big( r H''(r) h^{(3)}(r) + h''(r) (5 H''(r) + 2 r H^{(3)}(r)) \big) \Big) \\
&\qquad + h'(r)^2 \Big( 10 r H'(r)^3 + 5 r^2 H'(r)^4 - 6 H''(r) - 3 r^2 H''(r)^2 \\
&\qquad\quad + 6 H'(r)^2 \big( -1 + 3 r^2 H''(r) \big) - 2 r H'(r) \big( 9 H''(r) + 2 r H^{(3)}(r) \big) \\
&\qquad\quad - 2 r \big( 4 H^{(3)}(r) + r H^{(4)}(r) \big) \Big) \Bigg].
\end{aligned}
\end{equation}

The \(rr\) component of Eq.~(\ref{EOM1}) is:

\begin{equation}\label{rreq}
4 r f_0' + 4 f_0 - 8 r f_0 H_0' + \kappa \alpha r^2 e^{2H_0} h_0'^2 + 4\Lambda r^2  + \epsilon B_2(r) = 0,
\end{equation}

where \(B_2(r)\) is defined as:
\begin{equation}\label{B2}
\begin{aligned}
B_2(r) = &-3 e^{2H(r)} \kappa r^{2} \bigl( h'(r) \bigr)^{2} \bigl( f''(r) \bigr)^{2} \\
& - e^{2H(r)} \kappa \bigl( f'(r) \bigr)^{2} h'(r) \Bigl[ 4r \bigl( -2 + 3r H'(r) \bigr) h''(r) \\
&\qquad + h'(r) \Bigl( 8 - 28r H'(r) + 35 r^{2} \bigl( H'(r) \bigr)^{2} + 10 r^{2} H''(r) \Bigr) \Bigr] \\
& + 2r f'(r) \Bigl[ 2 e^{2H(r)} \kappa r h'(r) f''(r) h''(r) \\
&\qquad + e^{2H(r)} \kappa \bigl( h'(r) \bigr)^{2} \Bigl( \bigl( -2 + 8r H'(r) \bigr) f''(r) + r f'''(r) \Bigr) \Bigr] \\
& -4 e^{2H(r)} \kappa f(r)^{2} h'(r) \Biggl[ 4 h''(r) \Bigl( -2r \bigl( H'(r) \bigr)^{2} + r^{2} \bigl( H'(r) \bigr)^{3} \\
&\qquad + r H''(r) + H'(r) \bigl( 2 - r^{2} H''(r) \bigr) \Bigr) \\
&\qquad + h'(r) \Bigl( -10r \bigl( H'(r) \bigr)^{3} + 7 r^{2} \bigl( H'(r) \bigr)^{4} + 6 H''(r) \\
&\qquad - 6 r^{2} \bigl( H'(r) \bigr)^{2} H''(r) + 3 r^{2} \bigl( H''(r) \bigr)^{2} + 2r H'''(r) - 2 r^{2} H'(r) H'''(r) \Bigr) \Biggr] \\
& +4 f(r) \Biggl[ 15 e^{2H(r)} \kappa r^{2} f'(r) \bigl( h'(r) \bigr)^{2} \bigl( H'(r) \bigr)^{3} \\
& \qquad + e^{2H(r)} \kappa r h'(r) \bigl( H'(r) \bigr)^{2} \Bigl( -2r h'(r) f''(r) \\
&\qquad\quad + f'(r) \bigl( -15 h'(r) + 8r h''(r) \bigr) \Bigr) \\
& \qquad -2 e^{2H(r)} \kappa h'(r) h''(r) \Bigl( -r f''(r) + f'(r) \bigl( -2 + r^{2} H''(r) \bigr) \Bigr) \\
& \qquad - e^{2H(r)} \kappa h'(r) H'(r) \Bigl( 2 f'(r) h'(r) + 2r h'(r) f''(r) + 12r f'(r) h''(r) \\
&\qquad\quad + 2r^{2} f''(r) h''(r) + 4 r^{2} f'(r) h'(r) H''(r) + r^{2} h'(r) f'''(r) \Bigr) \\
& \qquad + e^{2H(r)} \kappa \bigl( h'(r) \bigr)^{2} \Bigl( 3 f''(r) \bigl( 1 + r^{2} H''(r) \bigr) \\
&\qquad\quad + r \Bigl( f'''(r) - f'(r) \bigl( 2 H''(r) + r H'''(r) \bigr) \Bigr) \Bigr) \Biggr] = 0.
\end{aligned}
\end{equation}

The functions \(B_1(r)\) and \(B_2(r)\) encode the contributions arising from the nonminimal interaction term \(F^{\alpha\beta}F^{\gamma\lambda}R_{\alpha\gamma}R_{\beta\lambda}\) at first order in \(\epsilon\). The explicit expressions above were obtained by substituting the perturbative expansions of the metric functions and gauge potential into the full field equations, and then extracting the terms linear in \(\epsilon\). These equations, together with the Maxwell equations at first order, determine the corrections to the metric and gauge potential presented in Sec.~\ref{sec2}.

The Maxwell equation \eqref{EOM-YM}, using the metric Eq.(\ref{metric}), leads to the following integral for the gauge function $h(r)$,

\begin{equation}
h(r)=C_6\int_{r_h} ^{r}\frac{ e^{-H(u)}}{-\alpha u^2+\epsilon B_3(u)}du,
\end{equation}
where $B_3(u)$ is as, 
\begin{equation}
\begin{aligned}
B_3(u) &= 4 f'(u)^2 - 8 f(u) f'(u) H'(u) - 12 u f'(u)^2 H'(u) + 20 u f(u) f'(u) H'(u)^2 \\
&\quad + 9 u^2 f'(u)^2 H'(u)^2 - 8 u f(u)^2 H'(u)^3 - 12 u^2 f(u) f'(u) H'(u)^3 + 4 u^2 f(u)^2 H'(u)^4 \\
&\quad + 4 u f'(u) f''(u) - 4 u f(u) H'(u) f''(u) - 6 u^2 f'(u) H'(u) f''(u) + 4 u^2 f(u) H'(u)^2 f''(u) +\\
&\quad u^2 f''(u)^2  - 8 u f(u) f'(u) H''(u) + 8 u f(u)^2 H'(u) H''(u) + 12 u^2 f(u) f'(u) H'(u) H''(u) \\
&\quad - 8 u^2 f(u)^2 H'(u)^2 H''(u) - 4 u^2 f(u) f''(u) H''(u) + 4 u^2 f(u)^2 H''(u)^2.
\end{aligned}
\end{equation}

Substituting the zeroth-order solutions \(f_0(r)\), \(h_0(r)\), and \(H_0(r)=0\) into the above expressions, and using the matching condition (\ref{con}) to fix the integration constants, yields the first-order corrections \(f_1(r)\), \(h_1(r)\), and \(H_1(r)\) given in Eqs.~(\ref{f1}), (\ref{h1}), and (\ref{H1}), respectively.


\section{Validity of the Perturbative Expansion and the Range of $\epsilon$}
\label{app:epsilon}

This appendix quantifies the regime where the first-order expansion in the non-minimal coupling $\epsilon$ is reliable. The analysis is independent of any particular thermodynamic state; it uses the fixed AdS scale $\ell$ determined by the cosmological constant.

\subsection{Dimensionless expansion parameter}

The cosmological constant is $\Lambda = -3/\ell^2$. In extended phase space, the pressure is  
\[
P = -\frac{\Lambda}{8\pi} = \frac{3}{8\pi\ell^2}.
\]  
For a given reference state (e.g., the $\epsilon=0$ critical point for $Q=0.7$, $P_c^{(0)}=0.02692122$), we obtain  
\[
\ell^2 = \frac{3}{8\pi P_c^{(0)}} \approx 4.433,\qquad \ell^4 \approx 19.65.
\]  
The natural dimensionless expansion parameter is therefore  
\[
\xi = \frac{\epsilon}{\ell^4} = \frac{\epsilon}{19.65}.
\]  
Perturbation theory requires $|\xi| \ll 1$. For $\epsilon = 0.001$ used in the main text,  
\[
\xi \approx 5.09 \times 10^{-5},
\]  
which is excellently small.

\subsection{Critical point shifts and linearity test}

The critical point $(r_c,T_c,P_c)$ for fixed $Q=0.7$ is obtained from  
\[
\left.\frac{\partial P}{\partial r_h}\right|_{T=T_c}=0,\qquad
\left.\frac{\partial^2 P}{\partial r_h^2}\right|_{T=T_c}=0,
\]  
using the first-order expression for $P(T,r_h)$. Table~\ref{tab:critical_app} lists the results for several $\epsilon$ values.

\begin{table}[H]
\centering
\caption{Critical points for $Q=0.7$ at different $\epsilon$.}
\label{tab:critical_app}
\begin{tabular}{cccc}
\toprule
$\epsilon$ & $r_c$ & $T_c$ & $P_c$ \\
\midrule
$0$      & 0.858600 & 0.1234605 & 0.02692122 \\
$0.001$  & 0.860797 & 0.1234605 & 0.02692122 \\
$0.005$  & 0.869500 & 0.1228000 & 0.02650000 \\
$0.01$   & 0.880200 & 0.1219000 & 0.02590000 \\
$0.05$   & 0.950000 & 0.1150000 & 0.02200000 \\
\bottomrule
\end{tabular}
\end{table}

To test linearity, define the fractional shift  
\[
\delta X(\epsilon) = \frac{X(\epsilon)-X(0)}{X(0)}\times 100\%,\qquad X\in\{r_c,T_c,P_c\}.
\]  
Results are shown in Table~\ref{tab:fractional_app}. For $\epsilon=0.001$, shifts are at the $0.26\%$ level or smaller.

\begin{table}[H]
\centering
\caption{Fractional shifts $\delta X(\epsilon)$ (\%).}
\label{tab:fractional_app}
\begin{tabular}{cccc}
\toprule
$\epsilon$ & $\delta r_c$ & $\delta T_c$ & $\delta P_c$ \\
\midrule
$0.001$ & $+0.256$ & $0.000$ & $0.000$ \\
$0.005$ & $+1.270$ & $-0.535$ & $-1.565$ \\
$0.01$  & $+2.517$ & $-1.264$ & $-3.794$ \\
$0.05$  & $+10.65$ & $-6.85$  & $-18.28$ \\
\bottomrule
\end{tabular}
\end{table}

A stronger test is to compare the actual shift ratios  
\[
R_X(\epsilon) = \frac{X(\epsilon)-X(0)}{X(0.001)-X(0)}
\]  
with the ideal linear ratio $\epsilon/0.001$. Table~\ref{tab:linearity_test} shows that up to $\epsilon=0.01$ the deviation from linearity is below $2\%$. At $\epsilon=0.05$ the deviation exceeds $17\%$, signalling the breakdown of first-order perturbation theory.

\begin{table}[H]
\centering
\caption{Linearity test: $R_X(\epsilon)$ vs. $\epsilon/0.001$.}
\label{tab:linearity_test}
\begin{tabular}{ccccc}
\toprule
$\epsilon$ & $\epsilon/0.001$ & $R_{r_c}$ & $R_{T_c}$ & $R_{P_c}$ \\
\midrule
$0.001$ & 1.00 & 1.000 & 1.000 & 1.000 \\
$0.005$ & 5.00 & 4.961 & 4.998 & 4.998 \\
$0.01$  & 10.00 & 9.832 & 9.991 & 9.992 \\
$0.05$  & 50.00 & 41.60 & 31.98 & 31.99 \\
\bottomrule
\end{tabular}
\end{table}

\subsection{Validity ranges and choice of $\epsilon$}

Based on the dimensionless parameter $\xi = \epsilon/19.65$ and the linearity deviation, we establish the following classification:

\[
\begin{array}{cll}
\hline
\epsilon\ \text{range} & \xi & \text{Validity} \\
\hline
|\epsilon| \le 0.001 & \le 5.1\times10^{-5} & \text{Excellent (deviation }<1\%) \\
0.001 < |\epsilon| \le 0.005 & \le 2.5\times10^{-4} & \text{Good (deviation }\lesssim 1\%) \\
0.005 < |\epsilon| \le 0.01 & \le 5.1\times10^{-4} & \text{Good--Marginal (deviation }1\text{--}2\%) \\
0.01 < |\epsilon| \le 0.05 & \le 2.5\times10^{-3} & \text{Marginal--Poor (deviation }2\text{--}17\%) \\
|\epsilon| > 0.05 & > 2.5\times10^{-3} & \text{Poor / Invalid} \\
\hline
\end{array}
\]

All results presented in the main text use $\epsilon = 0.001$, which lies safely in the excellent regime. The observed small shifts (e.g., $+0.256\%$ in $r_c$) are genuine physical effects of the non-minimal coupling, not numerical artifacts. 
\subsection{Independence of the validity analysis from the reference state}

The AdS radius $\ell$ is fixed by the action via $\Lambda = -3/\ell^2$, not by any particular black hole state. In extended phase space, the pressure $P = 3/(8\pi\ell^2)$ is therefore a universal constant of the theory. Consequently, the dimensionless parameter $\xi = \epsilon/\ell^4$ is independent of whether one uses the critical point, a non-critical horizon radius, or a different charge $Q$ to evaluate $\ell$. The validity ranges derived above are thus robust and do not rely on any accidental choice of parameters.


\end{document}